\documentclass[11pt,preprint]{aastex}

\newcommand\be{\begin{equation}}
\newcommand\en{\end{equation}}

\usepackage{amsmath}
\shorttitle{Accretion in TO}
\shortauthors{Sicilia-Aguilar et al.}

\begin{document}

\title{Accretion in Evolved and Transitional Disks in Cep~OB2: Looking for the Origin of the Inner Holes}

\author{Aurora Sicilia-Aguilar\altaffilmark{1}, Thomas Henning\altaffilmark{1}, Lee W. Hartmann\altaffilmark{2}}

\altaffiltext{1}{Max-Planck-Institut f\"{u}r Astronomie, K\"{o}nigstuhl 17, 69117 Heidelberg, Germany}
\altaffiltext{2}{University of Michigan, 830 Dennison 500 Church St., Ann Arbor, MI 48109}

\email{sicilia@mpia.de}

\begin{abstract}

We present accretion rates for a large number of solar-type stars in 
the Cep~OB2 region, based on U band observations. Our study comprises 
95 members of the $\sim$4 Myr-old cluster Tr~37 (including 
20 ``transition'' objects; TO), as well as the only CTTS in the
$\sim$12 Myr-old cluster NGC~7160. The stars show different disk 
morphologies, with the majority of them having evolved and flattened disks. 
The typical accretion rates are about one order of magnitude lower than
in regions aged 1-2 Myr, and we find no strong correlation between 
disk morphology and accretion rates. Although half of the TO are not
accreting, the median accretion rates of normal CTTS and accreting 
``transition'' disks are similar
($\sim$3$\times$10$^{-9}$ and 2$\times$10$^{-9}~$M$_\odot$~yr$^{-1}$, respectively).
Comparison with other regions suggests that the TO observed at different
ages do not necessarily represent the same type of objects, which is
consistent with the fact that the different processes that can lead to
reduced IR excess/inner disk clearing (e.g., binarity, dust coagulation/settling, photoevaporation,
giant planet formation) do not operate on the same timescales. Accreting TO in
Tr~37 are probably suffering strong dust coagulation/settling. Regarding the 
equally large number of non-accreting TO in the region, other processes, like 
photoevaporation, the presence of stellar/substellar companions, and/or giant 
planet formation may account for their ``transitional'' SEDs and negligible 
accretion rates.
\end{abstract}


\section{Introduction \label{intro}}

In the last years, $Spitzer$ observations
have repeatedly shown that disk dissipation is a complex process,
and that substantial evolution occurs to the dust grains and to the
SED structure before the disk is rendered optically thin and disappears (e.g. van Boekel 
et al. 2003; Apai et al. 2005; Lada et al. 2006; Sicilia-Aguilar et al. 2006a, 
2007, 2008b, 2009; Rodmann et al. 2006; Currie et al. 2009).
By the age of 4 Myr, the typical spectral
energy distribution (SED) shows that grain growth, dust settling,
or both, have contributed to decrease the flaring of the disk and/or to lessen
the dust opacity, resulting in lower luminosities in the near- and mid-IR compared 
to younger disks (Sicilia-Aguilar et al. 2006a). In some objects, the dust
evolution has gone so far, that the disks seem to have been caught in the 
final act of dispersion, being in an intermediate stage between a normal, Class II
object and a Class III source with an optically thin disk. Unlike the normal disks 
around classical T Tauri stars (CTTS), these ``transition'' objects (TO) show no 
significant near-IR excess, 
but retain optically thick disks at longer wavelengths,
which suggest that they have inner opacity holes and are suffering inside-out
clearing (Forrest et al. 2004; D'Alessio et al. 2005; McCabe et al. 2006; 
Sicilia-Aguilar et al. 2006a). How the dusty disk dissipates, and whether 
all disks go through the TO phase before disappearing, remains uncertain.
Regions in the 1-10 Myr age range are found to contain TO, although they seem to 
be more common in older clusters (Sicilia-Aguilar et al. 2006a, 2009; Currie et al. 2009).
The fact that at any age we find objects with and without disks, and
the abundance of flattened disks (optically thick, geometrically thin) 
in clusters of intermediate age (Sicilia-Aguilar et al. 2006a; Currie et al. 2009) 
suggest a variety of physical processes leading to disk dissipation.

In addition to dust, the disk contains about 99\% of gas, which controls the dust dynamics, 
at least until self-gravitating clumps have formed (Schr\"{a}pler \& Henning 2004; 
Johansen \& Klahr 2005; Johansen et al. 2007). The decrease in disk fraction
over time is accompanied by a ``parallel'' decrease in accretion activity and in the
magnitude of the accretion rates (Hartmann et al. 1998; Sicilia-Aguilar et al. 2006b;
Fedele et al. 2009). Nevertheless, detecting the gas in the 
inner few-AU where planet formation is thought to occur, is complicated.
Near-IR spectroscopy has revealed the presence (or lack)
of gas in the innermost disk of some T Tauri stars with different disk
morphologies (Najita et al. 1996, 2003; 
Carr et al. 2001; Rettig et al. 2005; Pascucci et al. 2006, 2009; Goto et al. 2006;
Carmona et al. 2007, 2008; Salyk et al. 2009),
but the procedure is too expensive to include large, 
statistically significant samples of disks. Detecting the gas as it falls onto
the star via chromospheric accretion (K\"{o}nigl et al. 1991) is an easier
procedure to constrain the presence of gas in the innermost disk for large
number of objects. 

Objects that have partially 
dispersed/coagulated their disks (TO) are especially important to understand 
the factors involved in disk evolution. While there is no consensus about the
definition and characteristics of TO, we here follow the rule of considering
TO those objects that show zero or negligible IR excess at $\lambda <$6$\mu$m and 
normal IR excess (CTTS-like) at longer wavelengths, compared
to the photospheric emission. According to our definition, TO have 
reduced IR colors with [3.6]-[4.5]$<$0.2 mag and a sharp increase in the excesses
usually around 8$\mu$m (Sicilia-Aguilar et al. 2006a. A similar definition
has been used by Muzerolle et al. (2009). For solar-type (late G-early M\footnote{A 1M$_\odot$
star aged $\sim$1 Myr has an spectral type $\sim$K6 (Siess et al. 2001).}) stars, this
suggests the presence of an inner hole or a highly evolved/settled and probably optically 
thin inner disk with radius of a few AU within a disk that is flared and 
optically thick at larger distances from the star (D'Alessio et al. 2005), 
consistent with inside-out clearing of the disk (Hayashi et al. 1985).
The presence of accretion and the accretion rates are necessary 
parameters to distinguish between the different scenarios that create the inner 
holes of  the TO (photoevaporation, dust coagulation, planetesimal formation, 
formation of giant planets). Systems empirically classified as TO do not 
form a uniform class of objects, but include binaries (Ireland \& Kraus 2008), 
objects with giant planets (Setiawan et al. 2008), and cases where strong dust
coagulation and/or photoevaporation have produced an optically thin disk
(Pascucci \& Sterzik 2009). Since these different processes operate on
different timescales, it is not evident that the bulk of TO in regions 
with different ages should be dominated by the same type of objects
(Alexander \& Armitage 2009).

Here we present the results of a deep U band survey of the young clusters 
Tr~37 and NGC~7160, both part of the Cep~OB2 region. Section \ref{data}
describes the observations, data reduction, and the calculation of the 
accretion rates. In Section \ref{analysis} we analyze the 
potential correlations of \.{M} with disk structure, age, and stellar mass, and
explore the fraction of stars that display variable accretion rates.
Finally, we summarize our conclusions in Section \ref{conclu}. 

\section{Observations, data reduction, and accretion rates \label{data}}

\subsection{The Cep~OB2 region: previous observations \label{old-obs}}

Cep~OB2 is a bubble-shaped star-forming region (Patel et al. 1998) located at 
900 pc distance (Contreras et al. 2002). It contains several OB stars and the
young clusters NGC~7160 (median age $\sim$10-12 Myr) and Tr~37 (median age $\sim$4 Myr). 
Tr~37 and NGC~7160 contain a significant population of 
solar-type stars ($\sim$180 and $\sim$60 late G-M2-type members, respectively). 
A younger population (aged $\sim$1 Myr) is located to the west of Tr~37 
(Sicilia-Aguilar et al. 2005b, 2006a). Optical spectroscopy
was used to determine the membership, spectral types, and the presence of accretion. 
The clusters have also been observed with $Spitzer$ IRAC ($\lambda$=3.6-8.0 $\mu$m) 
and MIPS ($\lambda$=24 and 70 $\mu$m; Sicilia-Aguilar et al. 2006a). In Tr~37,
the disk fraction for solar-type stars is $\sim$45\%, and more than 95\% 
of the disks show low IRAC fluxes (compared to Taurus), which suggests dust 
evolution (grain growth/settling), especially strong in the innermost disk. 
About 10\% of the objects are TO. In NGC~7160, 
only two of the known 60 solar-type members conserve their disks.
The accretion rates of 45 members of Tr~37 had been estimated via U band photometry
(Sicilia-Aguilar et al. 2005b). UVRI observations were obtained with the 1.2m telescope
at the the Fred Lawrence Whipple  Observatory (FLWO), but
the VRI and U band data were not taken simultaneously, and the survey
was complete only down to few times 10$^{-9}~$M$_\odot$~yr$^{-1}$. 
High-resolution H$\alpha$ spectroscopy of the Tr~37 members
(Sicilia-Aguilar et al. 2006b) revealed many more accreting objects,
suggesting that deeper U band observations were necessary to
fully study the dependence of accretion on disk structure, age, and
stellar mass.

\subsection{Observations and data reduction}

The U~V~R$_J$~I$_J$ observations of Tr~37 were obtained in service mode during 
3 nights in 2007 June 9-11, using the wide field camera LAICA mounted on the 3.5~m 
telescope in Calar Alto, Spain. LAICA is a 2$\times$2 mosaic of 4 CCDs 
(which we name a,b,c,d), each covering a 15.3'$\times$15.3' field of view with 
a large gap (15.3'$\times$15.3') in between. We obtained four pointings in 
Tr~37 (named 1,2,3,4 in the Table \ref{obs-table}), in order to cover a large 
field (approximately 45'$\times$45', containing most of the Tr~37 members) and
to avoid the brightest stars in the cluster. The exposure time was 3$\times$120s
in U, and for V~R$_J$~I$_J$, we obtained short and long exposures 3$\times$10s and
3$\times$60s, respectively. The 3 dithered exposures ensure a good removal of
cosmic rays, minimize CCD defects, and provide
a large dynamical range, complete in the magnitude ranges U$\sim$15-21, 
V$\sim$13-21, R$_J$$\sim$12-21, and I$_J$$\sim$11-20 (from now on, R$_J$ and
I$_J$ are simply written R and I). The time span of the observations 
was less than 1.5~h for the objects observed in the same night. Pointing 1 (UVRI) 
was observed during the first night. During the second night, we observed the 
pointing 2 (UVRI) and U band for the pointing 3. During the third night, we obtained 
VRI for the pointing 3, and UVRI for the pointing 4. Therefore, all the UVRI 
observations for the pointings 1,2, and 4 can be considered nearly-simultaneous, 
and the U band observations of pointing 3 are separated by $\sim$23~h from
the pointing 3 VRI (see Table \ref{obs-table}).
All the nights were photometric, with seeing in the range 0.6-1.3".

The observations of NGC~7160 were obtained in 2007 October 5, using the 70 cm 
King telescope of the Max-Planck-Institut f\"{u}r Astronomie at the K\"{o}nigstuhl,
Heidelberg, Germany. The telescope is equipped with a CCD with a field of view 
$\sim$18'$\times$18' and standard UBVR$_J$I$_J$ filters. We followed a similar 
strategy, obtaining 3 exposures of 10s and 240s in U, and 3$\times$120s for the rest of
filters, in a field centered on the cluster, which includes the only two low-mass sources  
with disks detected with $Spitzer$ in NGC~7160 (only one of them was detected in U). 
Given the smaller diameter of this telescope, the survey
is complete down to U$\sim$18, V$\sim$18, R$\sim$17.5, I$\sim$17 only.
The night was clear but not photometric, and the seeing was poor (2"). 

The data from the two telescopes were reduced following standard procedures with 
IRAF\footnote{IRAF is distributed by the National Optical Astronomy Observatories,
which are operated by the Association of Universities for Research
in Astronomy, Inc., under cooperative agreement with the National
Science Foundation.} within the \textit{noao.imred.ccdred} package to do the bias and flat 
field corrections, and \textit{noao.digiphot.apphot} for aperture photometry.
The flux calibration was obtained as relative (or differential) photometry
(Herbst et al. 2000; Brice\~{n}o et al. 2001), which we have also used 
to calibrate the variable star GM Cep (Sicilia-Aguilar et al. 2008a). For each
field and band, we matched our instrumental photometry to existing calibrated data.
The fact that most stars in the fields are non-variable allows us to derive the
calibration zeropoints for our data. The UVRI
data for Tr~37 and the VRI data for NGC~7160
were compared with our previous optical data from the Fred Lawrence Whipple Observatory
(Sicilia-Aguilar et al. 2005b), properly transformed from the Cousins FLWO filters into
the standard Johnsons system (Fernie 1983). Lacking U data for NGC~7160 from our FLWO 
catalog, the U band observations of NGC~7160 were calibrated comparing with the catalog
from De Graeve (1983). Since the calibration involves a large number of
stars observed simultaneously in the same CCD, the zeropoint uncertainties are typically
lower than 1\%, so the magnitude errors are dominated by the signal to noise 
ratio (S/N). The magnitudes and their errors are listed in Table \ref{data-table}.
To ensure the quality of the photometry, all images were visually 
inspected to remove all objects that could be affected by 
nearby stars, scattered light from bright sources, or CCD artifacts. 
We also compared the photometry from our previous studies (Sicilia-Aguilar et al. 2004, 2005b)
and those stars displaying magnitude variations $\Delta$V$\geq$0.5 mag
are labeled as ``variables" in Table \ref{data-table}. Note that $\sim$65\%
(62/96) of the objects display small magnitude variations ($\geq$0.1 mag; see Section \ref{variability}).

\subsection{Accretion rates \label{accretion}}

The accretion rates were derived from U broad band photometry
following the correlation between accretion luminosity (L$_{acc}$) and
U band excess luminosity (L$_U$) found by Gullbring et al. (1998):

\begin{equation}
\log(L_{acc}/L_\odot) = 1.09 \log(L_U / L_\odot) + 0.98 \label{lacc-eq}
\end{equation}

The accretion luminosity corresponds to the energy released by the matter falling 
onto the star. The accretion rate can be obtained from the accretion luminosity 
if the stellar mass and radius (M$_*$, R$_*$) are known, assuming that the accreted 
matter reaches the star in free fall via chromospheric accretion from a distance 
R$_{in}$$\sim$5~R$_*$: 

\begin{equation}
L_{acc} \sim G M_* \dot{M} / R_* (1-R_* / R_{in}) \label{mdot-eq}
\end{equation}

To calculate the excess of luminosity in U band (L$_U$) for a given object, we need 
to know the measured U band luminosity and the expected U band luminosity for a 
star with the same spectral type and luminosity. A good
extinction correction and accurate spectral type are essential. The spectral types
for the Tr~37 and NGC~7160 members were derived from low-resolution spectroscopy
in Sicilia-Aguilar et al. (2004) and Sicilia-Aguilar et al. (2005b), and are
accurate up to 1 subtype (for K and M stars) and 2 subtypes (for G-type stars). We 
calculated the extinction from the V-R and V-I colors, applying the relations in
E(V-R)=0.249 A$_V$ and E(V-I)=0.521 A$_V$ (Cardelli et al. 1989) and considering
the standard colors for each spectral type (Kenyon \& Hartmann 1995). Circumstellar 
matter may produce a non-standard extinction law, but the similar A$_V$ values 
obtained from V-R and V-I suggest that the main source of extinction towards 
Cep~OB2 is standard interstellar matter.  
Due to pointing offsets, a few sources lack V photometry, so
we derived their extinction from E(R-I)=0.272 A$_V$. Finally, for the objects
for which only one band (V, R, or I) was available, we consider previous measurements
of the extinction (Sicilia-Aguilar et al. 2004, 2005b) with an error
of 0.45 mag (corresponding to the standard deviation of A$_V$ in the cluster). 
If A$_V$ has not been measured previously, we take it to be 
the average of the cluster with its standard deviation (A$_V$=1.67$\pm$0.45 mag; 
Sicilia-Aguilar et al. 2005b).  The final 
uncertainties in the extinction reflect the errors in the photometry,
spectral type, and the variations between A$_V$ calculated via E(V-R) and E(V-I).

The U band luminosity is derived from the U magnitude corrected by the extinction. 
The U band luminosity due to the stellar photosphere is calculated using the U-V
and V-I colors and the bolometric correction for standard stars (Kenyon \& Hartmann 1995),
together with the extinction-corrected I magnitude (or, in objects without I magnitude,
R or V). Magnitudes are transformed into luminosities taking into account the 
zero point flux and bandwidth for U band 
(4.19$\times$10$^{-9}$ erg s$^{-1}$ cm$^{-2}$ \AA$^{-1}$ and 680~\AA, respectively). The 
stellar masses (M$_*$) and ages are calculated from the extinction-corrected 
V vs. V-I diagram and the Siess et al. (2000) isochrones, and the stellar radii (R$_*$)
are obtained from the total stellar luminosities and the effective temperatures. 
In the few cases where no V or no I are available, we use the total photospheric luminosity 
and effective temperature on a HR diagram and the Siess et al. (2000) isochrones 
to derive the stellar parameters. Knowing M$_*$, R$_*$  and the excess
U band luminosity L$_U$, the accretion luminosity and accretion rate are derived from 
Equations \ref{lacc-eq} and \ref{mdot-eq}.  The results are listed in Table \ref{acc-table}.

Given the number of parameters involved in the mass accretion calculation, 
it is important to assess if the measured \.{M} is significant.
Since the quantities required for the calculation of \.{M} are not independent 
(in particular, the two main contributors to the uncertainty, the spectral
type and the extinction, are strongly correlated), standard error propagation 
largely overestimates the errors. Moreover, the relation between the U magnitude
and L$_U$ is highly non-linear, so the errors are asymmetric in the positive and
negative directions. We therefore estimated the errors generating 5000 sets of artificial
data per object. The artificial data were constructed by adding Gaussian noise
with an amplitude corresponding to the measured errors to the involved parameters.
We then consistently derived A$_V$ and repeated the calculation of \.{M} for 
each one of the 5000 sets. The errors 
for each star are derived from the \.{M} distribution as the standard deviation
in the positive and negative directions. We consider an accretion rate to be safely 
detected ($>$0.99 probability) if L$_U$ is positive in more than 99\% of the 
simulations. If not, we estimate a 3-$\sigma$ upper limit to \.{M}, 
except for those objects for which H$\alpha$ observations have
ruled out the presence of accretion, especially if our $Spitzer$ data reveal no
excess emission, given that H$\alpha$ is a more sensitive criterion for low
accretion rates than L$_U$. For these narrow-H$\alpha$ objects, we consider the
accretion rate to be $<$10$^{-11}$~M$_\odot$~yr$^{-1}$, which is in agreement with
the line profile fitting for M-type stars (Muzerolle et al. 2003) and with the
observations of late-K objects in the $\eta$ Cha cluster (Lawson et al. 2004)\footnote{
Note that the limit will be lower for M-type objects and higher for the early K
and G stars.}. For simplicity, these ``non-accreting'' objects are labeled
with \.{M}=0~M$_\odot$~yr$^{-1}$ in  Table \ref{acc-table}.

In addition, there are other sources of error: uncertainties in the stellar
radius, stellar mass, and size of the stellar chromosphere. These uncertainties 
do not change the detectability of the accretion rate, but vary \.{M} 
by a typically small multiplicative factor (2-3; Gullbring et al. 1998). 
Since the accretion rates are found to vary by a factor 
of few when the same star is measured over time, 
these uncertainties do not affect our conclusions. 
The ability to detect accretion via U band excess emission also depends
on the spectral type, which in our sample varies between G8 and M2.5 (median
spectral type K7). For M stars, which have small intrinsic U band photospheric luminosity, 
small accretion rates result in detectable U band excess. 
For early K and G stars, a contrast problem appears as the U photospheric 
luminosity is much stronger than the L$_U$ caused by relatively 
large accretion rates (10$^{-8}$-10$^{-9}$~M$_\odot$~yr$^{-1}$). Therefore, some of 
our \.{M} measurements for the most massive stars are only (high) upper limits.
This has to be taken into account as a source of bias when 
studying the relation between the stellar mass and the accretion rate.
For the typical late K-early M star in our sample,
our U band photometry is sensitive to accretion rates of the order of
$\sim$10$^{-10}$~M$_\odot$~yr$^{-1}$. 

Comparing with our previous results, in particular, with the H$\alpha$ 
high-resolution spectroscopy with Hectochelle (Sicilia-Aguilar et al. 2006b), 
we find a very good agreement. All the objects with broad H$\alpha$ are 
found to be accreting or have upper limits consistent with the typical accretion
rates found in other cluster members.
We found an inconsistency between the spectral types in Sicilia-Aguilar et al.
(2005b) and the VRI colors of two G-type stars, 12-1091 and 11-581. They appeared
to be too red and had U magnitudes below the photospheric level, even
assuming the large extinctions derived from VRI. Considering that our spectral
types were more uncertain for G-type stars, we checked our spectra (taken
with FAST and Hectospec; Sicilia-Aguilar et al. 2005b) comparing them
to new standard stars and similar cluster members. The new spectral types
(K2 for 12-1091 instead of G2.5; G9 for 11-581 instead of G5) are fully
consistent with the photometry, and produce a more reliable value of
A$_V$ (with a similar value when derived from E(V-I) and E(V-R)).
Two objects are inconsistent in the sense that they lack evidence of a disk 
and have narrow H$\alpha$ profiles, but display U band excess (12-1984 and 12-1422). 
They both have non-simultaneous observations (pointing 3c), which may explain
the disagreement. The star 12-1422 is nevertheless a special case, since it
does not display any detectable IR excess nor H$\alpha$ broadening, but had a high H$\alpha$ EW
(-17 \AA\ for a M0 star) measured with our low-resolution spectra. Although its
accretion rate is close to our detection limit, we might consider it as a 
potential accreting TO with a large inner hole, to be confirmed with future observations.
Accretion in objects with no excess at 8$\mu$m seems rare, although here we
confirm the case of 13-819, which had been previously detected to have broad
H$\alpha$ despite the lack of excess at all IRAC wavelengths.

Finally, two TO with narrow H$\alpha$ appear now to be accreting (12-1009 and
72-875), and two more have upper limits consistent with accretion (14-197 and 12-595). 
For 12-1009, the data are non-simultaneous, and the star is known to have strong 
variability. The star 72-875, which is probably a single-lined spectroscopic binary (SB1)
and has strong MIPS excess but narrow H$\alpha$, is now found to have a U band excess. 
IRS spectroscopy (Sicilia-Aguilar et al. in prep.) reveals that it is a TO with
no excess at wavelengths shorter than $\sim$7$\mu$m, so the object seems to have
suffered a large increase in its accretion rate within a few years
(see Section \ref{variability}). The TO 14-197 is a similar mass spectroscopic binary,
and we thus believe it is most likely non-accreting. The star 12-595 shows anomalous colors
and extinction, being probably affected by scattering, non-standard extinction, 
and/or strong variability. Its U and VRI magnitudes are non-simultaneous and its LAICA V 
magnitude differ from the FLWO values by 1 magnitude, which suggests that the object
is strongly affected by variable extinction by circumstellar material. 
Although they are a minority, all the inconsistent 
or potentially inconsistent objects are excluded from the analysis and discussion below.

As an additional note, here we confirm once more (as stated in Sicilia-Aguilar et al. 2006b)
that even detailed criteria to distinguish accreting and non-accreting stars
based on H$\alpha$ equivalent width (EW) alone (e.g., White \& Basri 2003) may misclassify
a certain number of objects with low accretion rates or active WTTS. This is in general
not important for statistical purposes, especially in young clusters with typically 
high accretion rates. Nevertheless, it may become a problem in older regions like Tr~37
and NGC~7160 where low \.{M} is the rule. In our study, the 
White \& Basri (2003) criterion for H$\alpha$ EW
would misclassify 7 objects out of 96 (11-1209, 11-581, 12-1091, 13-819, 13-1048, 
13-1250, 22-1418, see Table \ref{acc-table}). Therefore, other observations (especially
high-resolution H$\alpha$ spectroscopy) are important 
to constrain the presence of accretion in the objects with the lowest rates.

\section{Analysis\label{analysis}}

\subsection{Gas accretion and dust evolution \label{parallel}}

\begin{figure}
\plotone{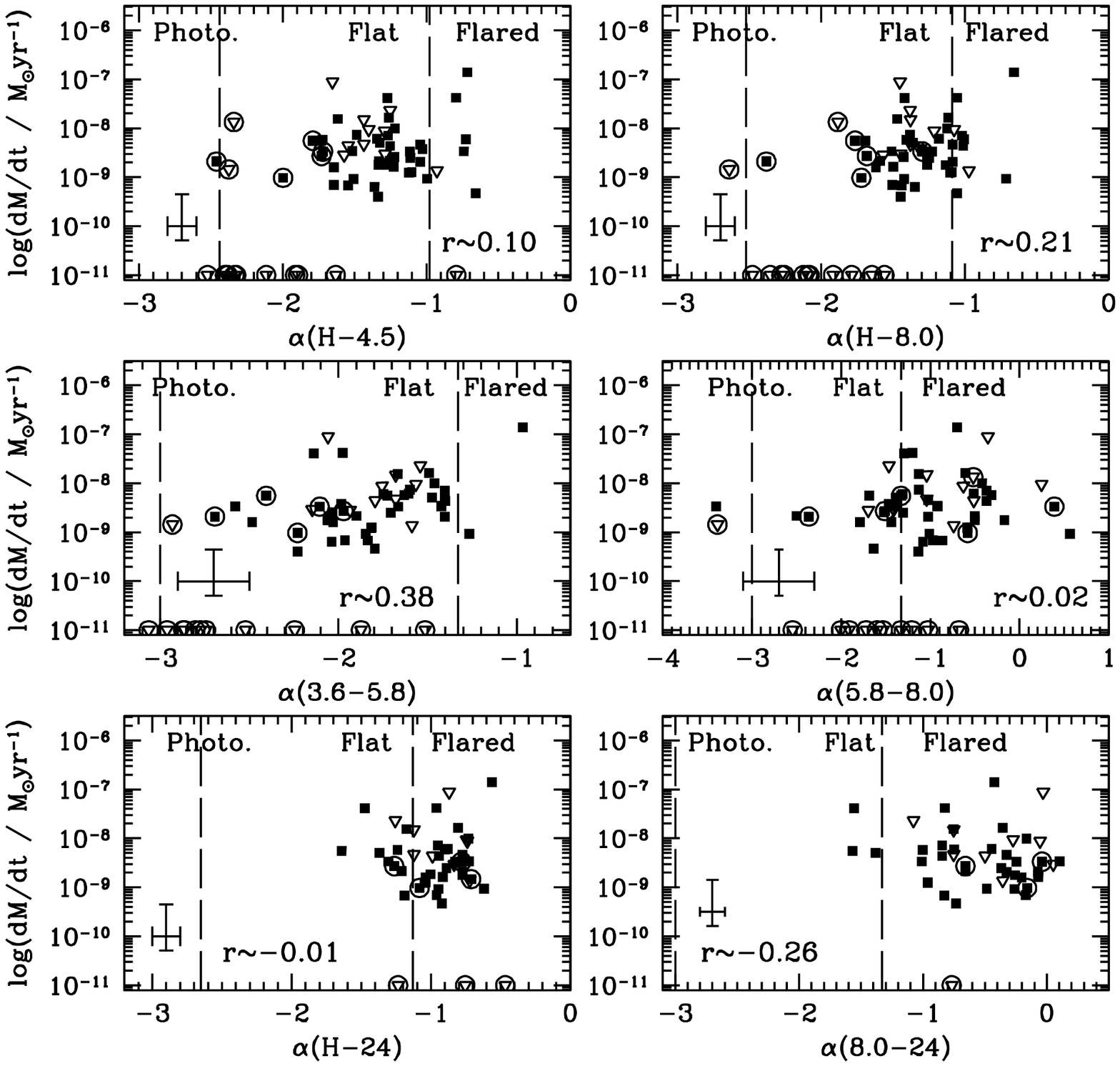}
\caption{Accretion rate versus the disk slope 
($\alpha$ = d log($\lambda$ F$_\lambda$)/d log($\lambda$)) at different wavelengths. 
The slopes of a stellar photosphere (for a K7 star) and an optically thick, geometrically 
thin disk (Kenyon \& Hartmann 1987) are displayed for comparison. Upper limits 
are shown as inverted open triangles. TO are marked with large circles. The typical error
bars, considering a 10\% uncertainty in the IRAC photometry, are also displayed.
To demonstrate the similarity of slopes between accreting and non-accreting TO, we also display
the $\alpha$ values of non-accreting TO, with upper limits to accretion 10$^{-11}$M$_\odot$ yr$^{-1}$
derived from H$\alpha$ spectroscopy. The generalized
``inside-out" evolution in Tr~37 (with nearly all objects showing stronger excess at
longer wavelengths; Sicilia-Aguilar et al. 2006a) is responsible for the 
changing $\alpha$ values at different wavelengths for each disk. 
\label{alpha-fig}}
\end{figure}

In order to check the relation between accretion rates
and dust evolution, we compare the disk morphology with the
measured \.{M}. It is not easy to define a tracer of disk morphology
or SED slope in the Cep~OB2 members, given that the typical SED slope
varies between the near- and mid-IR, with most objects showing a
``kink'' or slope change around 6-8 $\mu$m (Sicilia-Aguilar et al. 2006a).
Therefore, we measure the SED slope ($\alpha$) at different wavelengths,
and compare the $\alpha$ values with the measured accretion rates
(Figure \ref{alpha-fig}). According to our empirical definition,
TO have SED slopes consistent with photospheres at $\lambda <$6$\mu$m, together 
with negligible excess emission ([3.6]-[4.5]$<$0.2 mag) down to this wavelength. 
In our Spitzer study, we did not include as TO objects having only
a weak excess at 8$\mu$m (named as ``e8'' in Sicilia-Aguilar et al.
2006a) that were not detected at 24$\mu$m (due to our relatively high 
detection limits at 24$\mu$m). Nevertheless, our recent IRS spectroscopy 
has confirmed as TO the two TOe8 observed, which together with the facts
that no such objects were found in NGC~7160, and that
at least one is actively accreting (24-1796), suggested
that they are real TO. Summing up the number of ``e8'' TO, the TO fraction
in Tr~37 (including accreting and non-accreting TO) would be closer 
to $\sim$20\% of the total number of disks (or $\sim$10\% of the total
number of stars), and roughly half (9/20) of
the TO are accreting. 

Figure \ref{alpha-fig} does not reveal any strong trend of the accretion 
rate with the SED slope $\alpha$. The Spearman rank correlation coefficients 
(r) and probability of obtaining such r from randomly distributed 
data (p) for \.{M} versus $\alpha$ at different wavelengths vary between
0.01 and $\pm$0.3 (for r) and 0.97-0.02 (for p; see Table \ref{correlation-table}). 
There is a moderate correlation between
the slope $\alpha$ in the innermost disk (3.6-5.8$\mu$m) and the accretion
rates (r=0.38, p=0.02), which disappears at intermediate wavelengths, 
and becomes a weak anticorrelation (r=-0.26, p=0.12) for $\alpha$(8.0-24). 
The 24$\mu$m photometry was not deep enough to detect photospheres 
of solar-type stars at these wavelengths, but given the dispersion 
of the accretion rates for the objects detected and non-detected at 24$\mu$m,
we do not believe that the inclusion of objects with undetected 24$\mu$m 
fluxes would change the conclusions. This
suggests that, in general, objects with more evolved inner disks\footnote{Note
that TO and objects with ``kink'' SEDS have the largest (often positive) $\alpha$(8.0-24) slopes
in the sample.} tend to have smaller accretion rates. This correlation would
be expected if the changes in the dusty disk reflected the evolution of the total
surface density, with the dusty and gaseous components evolving ``in
parallel'' (Sicilia-Aguilar et al. 2006b; Fedele et al. 2009). Nevertheless, 
the fact that most of the Tr~37 disks are substantially
evolved and do not have large accretion rates make this trend weak. 
 
Therefore,
we find that the objects with the highest accretion rates tend to have 
the most flared disks, and that there are no TO with \.{M}$>$10$^{-8}$~M$_\odot$~yr$^{-1}$. 
Nevertheless, the distribution of accretion rates of TO and CTTS disks are not 
significantly different (within the limitations of small number statistics; 
see Section \ref{transition}), and the median accretion rate is the same 
for accreting TO and normal CTTS disks (2$\times$10$^{-9}$~M$_\odot$~yr$^{-1}$ and
3$\times$10$^{-9}$~M$_\odot$~yr$^{-1}$, respectively). 
Most of the CTTS in Tr~37 have relatively low accretion
rates (few times 10$^{-9}~$M$_\odot$~yr$^{-1}$), which makes them indistinguishable of
the accreting TO in terms of accretion rate. The large number of CTTS with
low \.{M} suggests that, in general, accretion rates
as low as 10$^{-9}$-10$^{-10}$~M$_\odot$~yr$^{-1}$ do not necessarily cause the
opening of an inner hole, as had been suggested by photoevaporation models
(Clarke et al. 2001; Ercolano et al. 2009b). 
 
Nevertheless, about half of the TO do not show U band excess nor H$\alpha$ line 
broadening, which sets upper limits to the accretion of $\leq$10$^{-11}~$M$_\odot$~yr$^{-1}$,
despite having similar disk slopes than the accreting TO (see Figure \ref{alpha-fig})
None of the objects with normal near-IR excess lacks evidence of accretion, which reveals an
important difference between TO and normal disks, and
points out relevant physical differences between accreting and
non-accreting TO, as has also been inferred from near-IR spectroscopy (Salyk et al. 2009).
Conversely, accretion in objects without any IR excess at $Spitzer$/IRAC 
wavelengths (and thus, large holes $>$10~AU) seems 
rare. We found one object (13-819) with broad H$\alpha$,
a measurable accretion rate, and no excess down to 8$\mu$m, although
two more objects could represent similar cases (21-895a and 12-1422). Since our MIPS 24~$\mu$m 
survey did not detect photospheres of solar-type stars, these objects
could still have an excess longwards of 8$\mu$m and thus a disk with a few-AU hole.

\subsection{Accretion evolution and the \.{M} vs. M relation \label{evolution}}

\begin{figure}
\plotone{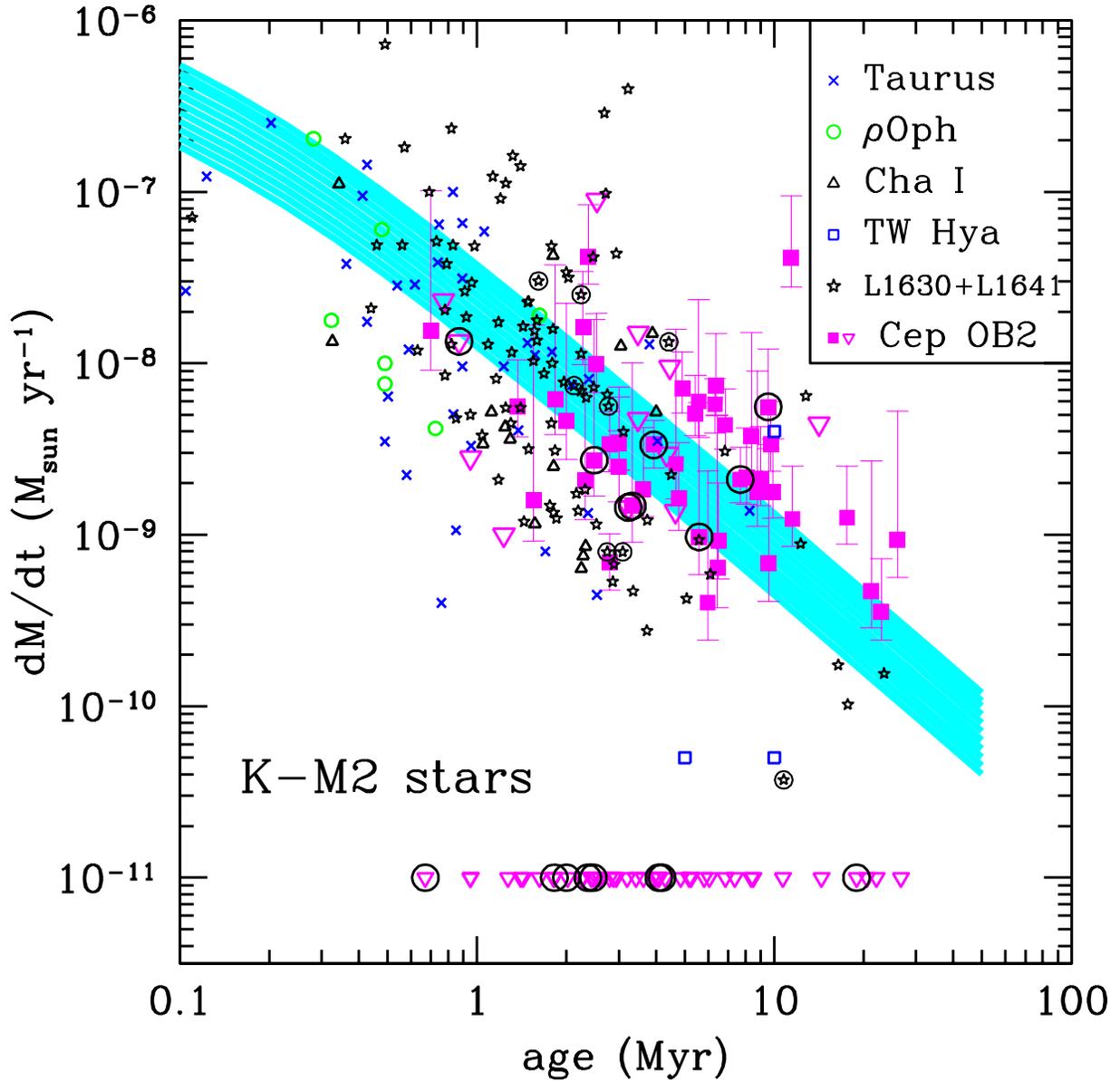}
\caption{Accretion versus stellar age. Detected accretion rates 
are represented by filled squares. Upper limits are 
shown as inverted open triangles. The upper limits based on H$\alpha$
spectroscopy (\.{M}$<$10$^{-11}$M$_\odot$ yr$^{-1}$ for narrow lines) are
also displayed. The 
plot includes data from other regions (Muzerolle et al. 2000; Fang et al. 2009). We
also display a collection of viscous disk evolutionary models (Hartmann et al. 1998)
for solar-type stars with initial disk masses $\sim$0.1-0.2 M$_\odot$,
constant viscosity $\alpha$=10$^{-2}$, and viscosity exponent $\gamma$=1. Note that
individual ages have large errors (not shown), mostly due to the presence of unresolved 
binaries and variability. TO in Cep~OB2 and in the Orion L~1630N and L~1641 clouds are marked
with a large circle around the corresponding symbol. The Spearman rank correlation coefficient
and probability of random distribution for the log(\.{M}) vs. log(age) relation 
are r$\sim$0.60, p$<$0.001 (including the
stars in all the regions, but excluding the non-accreting objects).
\label{agemdot-fig}}
\end{figure}

In the light of the new data for Cep~OB2, we examine the time evolution
of the accretion rate and the \.{M} vs M relation. Accretion rates for 
T Tauri disks are thought to decrease with time in a way that is consistent with
the evolution of a viscous disk (Hartmann et al. 1998; 
Muzerolle et al. 2000; Sicilia-Aguilar et al. 2005b, 2006b). 
Figure \ref{agemdot-fig} shows the accretion rates of the Cep~OB2 objects
versus their individual ages, compared to the values measured in 
other regions: Taurus, $\rho$ Ophiuchus, Chamaeleon I, and TW Hya
(Muzerolle et al. 2000), and the Orion L~1630N and L~1641 clouds 
(Fang et al. 2009), and to the predictions of viscous disk 
evolutionary models (Hartmann et al. 1998). Although the ages of individual 
objects are uncertain, given the presence of unresolved binaries and the 
stellar variability, the global age differences between regions
are real. For the G-type stars, the uncertainties in the 
birthline (Hartmann 2003) result in uncertain ages. Since G-type stars also 
have more massive disks and higher accretion rates, and our sample contains
very few stars earlier than K0 (the median spectral type is K7), we study 
the time variation of the accretion rate concentrating on objects with spectral 
types K-M2 only.

The Cep~OB2 data in Figure \ref{agemdot-fig} are consistent with the picture 
of viscous disk evolution, finding a moderate correlation between \.{M} and age 
(r$\sim$0.32, p=0.03). Although the ages of individual objects in Tr~37
are uncertain, the global age spread in Figure \ref{agemdot-fig} is real,
as there is a younger ($\sim$1 Myr) population in Tr~37, associated to the
IC~1396A globule, and an age gradient throughout the cluster (Sicilia-Aguilar
et al. 2005b). Including the rest of regions above mentioned, the
picture of viscous evolution becomes more clear, and the correlation between
age and \.{M} becomes very strong (r$\sim$0.60, p$<$0.001). 
The large range of \.{M} displayed by objects with a similar age confirms that
accretion (and disk) evolution does not happen in the same way in
all objects: At any given age, there are many objects that do not have disks and are
not accreting. In Tr~37, the number of non-accreting stars is similar
to the number of accretors. The initial disk mass and viscosity law affect
\.{M}~(t), but viscous evolution alone would not result in so many diskless
stars at intermediate ages. Photoevaporation may contribute to remove large 
amounts of gas and shorten the disk lifetimes (Clarke et al. 2001; 
Alexander et al. 2006b; Gorti et al. 2009).

Although the accretion data in Cep OB2 and other regions are roughly
consistent with viscous evolution (Figure \ref{agemdot-fig}), some differences 
observed provide clues about the different processes that result in disk 
evolution and dissipation, as well as the typical disk characteristics. The 
global distribution of accretion rates suggests that the initial disk mass,
which controls the initial \.{M} and thus results in higher or lower
accretion rates throughout the time, is higher than the 0.1~M$_\odot$
assumed by Hartmann et al. (1998)\footnote{Note that an initial
disk mass $\sim$0.1M$_\odot$ results in a disk mass comparable to the minimum
mass for the solar nebula, $\sim$0.01 M$_\odot$ at 1 Myr age.}. This disk mass had been 
calculated from millimeter observations, and we know now that they are probably
underestimated by up to an order of magnitude (Andrews \& Williams 2007).
An initially higher disk mass by a factor of 2-3 would thus be more consistent 
with the accretion rates we observe in Tr 37 (see Figure \ref{agemdot-fig}).
The mass would be even higher if we consider photoevaporation in
addition to viscous evolution, since up to 50\% of the initial disk
mass may be lost via photoevaporation (Gorti et al. 2009). In addition, the 
typical decay of \.{M} over time seems slower than previously assumed
in the models (\.{M}$\sim t^{-\eta}$, with $\eta \sim$1.2 rather than 1.5-2.8
as proposed by Hartmann et al. 1998). This is probably related to the fact
that after $\sim$4 Myr, only the most massive disks survive, while 
most of the others have been dissipated by a combination of viscous evolution 
and photoevaporation (Alexander et al. 2006b; Gorti et al. 2009). The slower 
decay we observe in our \.{M} versus age plot also suggests that the long-surviving 
disks may be biased towards certain radial viscosity laws. In fact, this
would be in good agreement with recent observations of Isella et al. (2009)
which also revealed a large variety in the viscosity parameter radial
exponent for different solar-type objects. 

\begin{figure}
\plotone{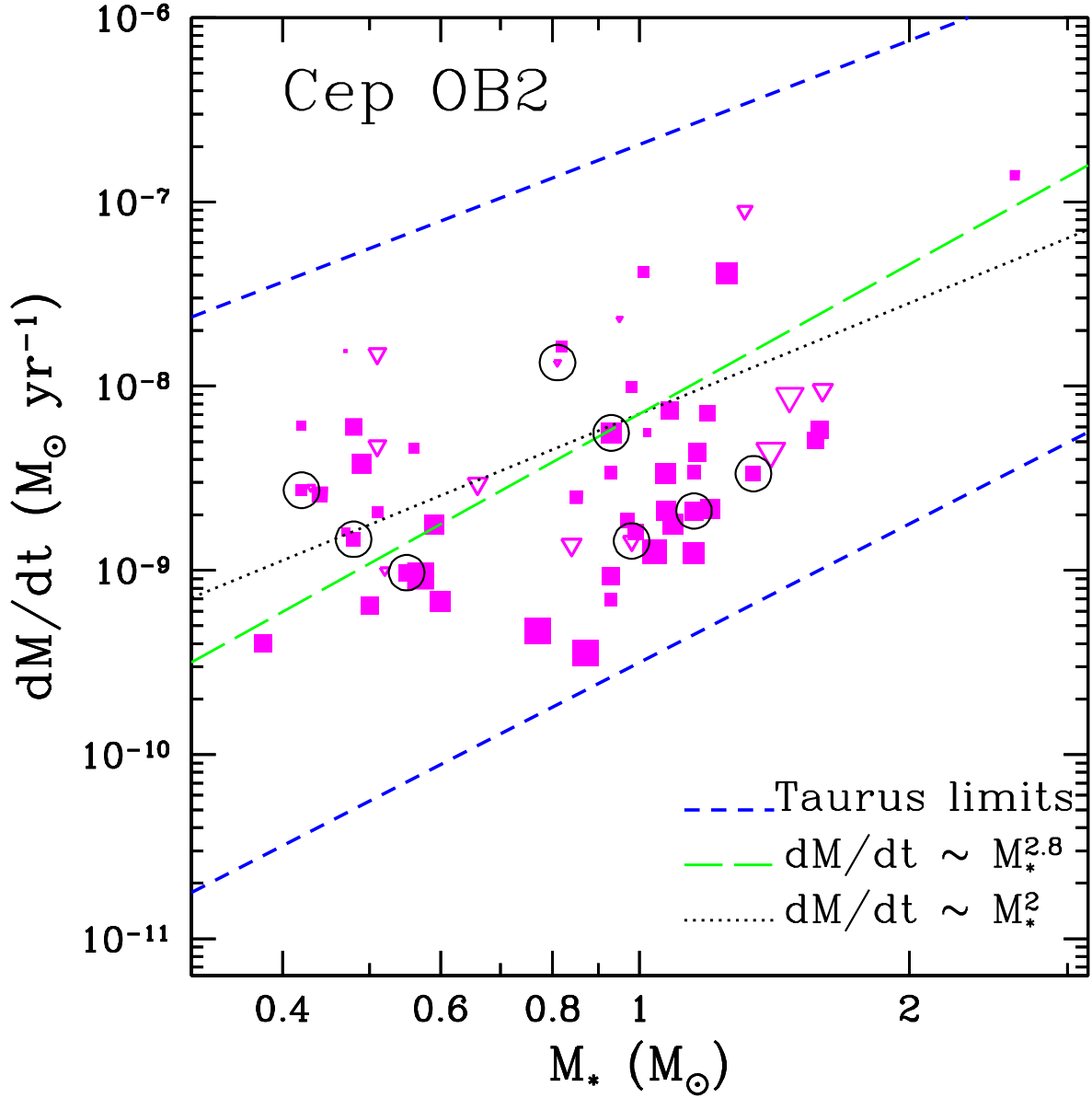}
\caption{Accretion versus stellar mass. Upper limits are shown as inverted open triangles.
The size of the symbols is proportional to the individual age (from $\sim$1 to 
$\sim$15 Myr). We show the limits of 
accretion and stellar mass for Taurus stars (short-dashed blue lines; 
Calvet et al. 2004), the slope for the \.{M}$\sim$M$_*^2$ relation (dotted black
line; Natta et al. 2004), and the slope for \.{M}$\sim$M$_*^{2.8}$ (long-dashed
green line; Fang et al. 2009). Note that these lines are merely indicative and not
fits to the data. The Spearman rank correlation coefficient and probability of
random distribution for the log(\.{M}) vs. log(M$_*$)
are r$\sim$0.32, p=0.03. \label{mmdot-fig}}
\end{figure}

Regarding the relation between the accretion rate and the stellar
mass (Figure \ref{mmdot-fig}), our results are consistent with
a moderate correlation between stellar mass and accretion rate (r$\sim$0.32, p=0.03)
with the slope range observed in Taurus (Calvet et al. 2004), although our
sample has systematically lower accretion rates. Our data are also
consistent with the \.{M}~$\propto$~M$_*^a$ relation, with $a\sim$2-3, as has
been suggested by Muzerolle et al. (2003), Natta et al. (2004) and, more recently, by
Fang et al. (2009), who found a larger index ($a\sim$3) for the lower-mass
stars. Nevertheless, the small mass range spanned by
the Cep~OB2 objects does not allow us to estimate the value of $a$, and the
fact that U band is less efficient detecting low accretion rates of
early K-G stars may also introduce an important bias.

\subsection{Accretion in TO: Constraining the origin of the ``inner holes''  \label{transition}}

TO and, in general, evolved disks, are a key factor to understand disk dissipation. 
Several mechanisms have been invoked to produce a TO disk, but the contribution of 
each one is not yet understood. The most important ones seem to be photoevaporation 
(Clarke et al. 2001; Alexander et al. 2006a,b, 2007; Gorti et al. 2009), 
dust coagulation (Ciesla 2007), and planet formation (Quillen et al. 2004; Edgar et al. 2007). 
Photoevaporation sets strong constraints to the lifetime and accretion rates of the disk, 
while strong dust coagulation and planet formation are more flexible.
In addition, the inner holes of some TO disks may not be related to pure disk evolution, 
but to the presence of close-in binaries (Ireland \& Kraus 2008). Although TO 
tend to be rare, summing up to 5-10\% of the total number of disks (Hartmann et al. 2005; 
Fang et al. 2009), some low-mass star forming regions like the Coronet cluster,
the MBM12 region, and the $\eta$ Cha cluster display very high number of TO and evolved
disks (Sicilia-Aguilar et al. 2008b; Meeus et al. 2009; Megeath et al. 2005;
Sicilia-Aguilar et al. 2009). Although the lack of near-IR excess in very low-mass
objects is not a clear sign of a clean inner hole, but may be also reproduced by
very flattened/settled disks around low luminosity stars (Ercolano et al. 2009a), the structural 
differences between these TO and the normal, flared disks are obvious and suggest
a strong degree of evolution. In older regions, the number of objects that seem to 
be ``in transition'' between Class II and Class III sources, including those with 
flattened/coagulated/settled inner disks, is higher than expected if the mechanisms leading 
to inside-out disk clearing operated in less than a Myr (Currie et al. 2009). 

\begin{figure}
\plotone{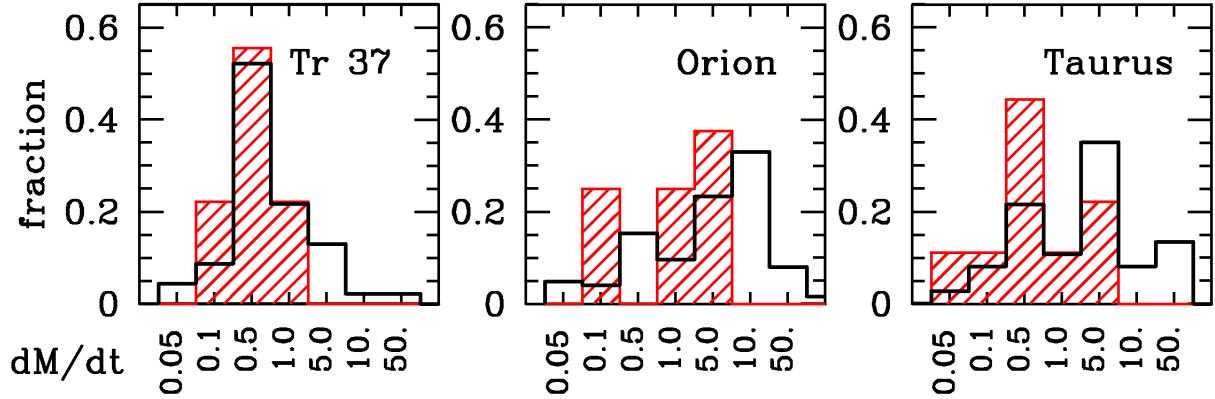}
\caption{Fraction of objects with accretion rates in a given range in
Tr 37, the Orion L~1630N and L~1641 clouds, and Taurus. Normal CTTS
are displayed in black, the dashed red histogram corresponds to the TO
in each region. The mass accretion rate is given in units of
10$^{-8}$M$_\odot$yr$^{-1}$. The fraction of objects with a given \.{M} range
is normalized to the total number of objects in each class for each 
region (CTTS and TO, respectively). \label{histo-fig}}
\end{figure}

In order to contrast the TO properties at different ages, we examined several well-known regions
at different stages of evolution, restricting ourselves to the range of solar-type stars 
(late G-M2). In Tr~37 the median accretion rate of the CTTS (based on 46 objects)
is  \.{M}= 3$\times$10$^{-9}$~M$_\odot$~yr$^{-1}$, which is not significantly
different from the median accretion rate of accreting TO (\.{M}= 2$\times$10$^{-9}$~M$_\odot$~yr$^{-1}$,
based on 9 objects). In Taurus, the median accretion rates of CTTS (based on 42
objects) and TO (9 objects) differ by one order of magnitude, being
\.{M}= 3$\times$10$^{-8}$~M$_\odot$~yr$^{-1}$ for CTTS and \.{M}= 3$\times$10$^{-9}$~M$_\odot$~yr$^{-1}$
for TO (Najita et al. 2007). In the L~1630N and L~1641 Orion clouds, Fang et al. (2009)
gives a similar accretion rate, \.{M}= 3$\times$10$^{-9}$~M$_\odot$~yr$^{-1}$, for both CTTS and TO
with spectral types G-M6, but if we restrict ourselves to the solar-type range,
we find median accretion rates of \.{M}= 1$\times$10$^{-8}$~M$_\odot$~yr$^{-1}$ for CTTS
and  \.{M}= 7$\times$10$^{-9}$~M$_\odot$~yr$^{-1}$ for TO (for 88 and 8 stars, 
respectively). Histograms of the accretion rates of CTTS and TO in the 
different regions are displayed in Figure \ref{histo-fig}.
Taurus and the Orion clouds have roughly the same ages (1-2 Myr), but the
differences between the accretion rate of TO and CTTS in Orion is rather
a factor $\lesssim$2 (thus similar to Tr~37, within small number statistics) 
than a factor 10. The accretion rates of TO in the work of Najita et al. 
may be underestimated, since they include upper limits to some TO that
are most likely not accreting (like CoKu Tau/4; D'Alessio et al. 2005;
Ireland \& Kraus 2008), while Fang et al. excludes the non-accreting TO
(\.{M}$<$10$^{-11}$~M$_\odot$~yr$^{-1}$) from the median, as we do here. 
The way \.{M} is derived is different in Fang et al. (2009),
who estimate the accretion rate from the H$\alpha$ luminosity,
compared to Najita et al. (2008), who use accretion rates derived from veiling and U band
excess, which are expected to match our work here. Nevertheless, the
H$\alpha$ luminosity vs. \.{M} calibration of Fang et al. (2009) was done 
using objects whose accretion had been measured from UV veiling and/or
U band excess (Gullbring et al. 1998; Dahm 2008; Herczeg \& Hillenbrand 2008), 
so we expect no systematic offsets when comparing the different datasets.

We find differences in the distribution of accretion rates of
normal CTTS and TO in the three regions (see Figure \ref{histo-fig}). The relative
numbers of TO that do not show accretion indicators are also different,
being higher in Tr~37 than in Orion ($\sim$50\% versus $\sim$30\%; 
Fang et al. 2009). In Tr 37, the fact that we find normal CTTS with low \.{M} 
and TO with similar accretion rates suggests that the difference in near-IR
excess for these two types of sources is related only to the dust
content (grain growth/dust settling), without significant differences in the gaseous 
disk component. In addition, the difference in the accretion rates of TO in regions with
different ages is not as striking as the evolution seen for CTTS 
(see Figures \ref{agemdot-fig} and \ref{histo-fig}).
This could be understood if the mechanisms producing the holes of TO
require a special range of accretion rates to be effective, or if the
changes produced in the disk structure of TO constrain the allowed range
of \.{M}. 

The order of magnitude difference between the accretion rates of 
CTTS and TO (for similar disk masses) observed by Najita et al. 
in Taurus is consistent 
with the opening of holes by Jovian planet formation. The 
predictions for the changes in the accretion rate with the presence of 
a planet suggest that the accretion through the gap is about 10\% of 
the accretion through the outer disk (Lubow \& D'Angelo 2006). Viscous
evolution would proceed relatively undisturbed after the formation of a planet
(Mordasini \& Klahr, private communication), and 
thus the accretion rates of TO would be expected to evolve with time as the accretion 
for normal CTTS disks does. Therefore, the lack of differences and \.{M} evolution seen 
in Tr~37 suggests that most of the accreting TO in this region are
not related to giant planet formation. Planet synthesis models (Mordasini
et al. 2009) predict that planet formation in disks with typical masses
would peak at an age $\sim$4~Myr, so we may expect a large number of
planet-related holes at the age of Tr~37. Nevertheless, the mass distribution
of observed planets (Bouchy et al. 2009) indicates that most of these objects 
would have masses insufficient to open a clean hole, or too large to allow 
accretion to continue. Therefore, we may expect that most of the disks 
harboring planets in Tr~37 would show either normal IR excess, or be
otherwise non-accreting TO.

The timescales for the onset of the different disk-clearing mechanisms
(close-in binaries, photoevaporation, strong grain coagulation/settling, and/or 
planetesimal/planet formation) are not the same, and the lifetimes of the resulting 
objects are also different. Therefore, it is not evident that the bulk of TO 
in regions with different ages ought to contain physically similar objects.
Rapid formation of a giant planet in a gravitationally unstable inner disk 
is nearly instantaneous (Boss 1997), although only a minority of
disks may be massive enough to suffer gravitational instability 
(Andrews \& Williams 2007). A hole due to the presence of
a binary companion would appear also immediately, but the resulting 
circumbinary disk is expected to survive less than a disk around a single 
star, with a lifetime $\sim$5 Myr (Bouwman et al. 2006). 
Photoevaporation (Clarke et al. 2001) is efficient only in combination with viscous evolution 
(Gorti et al. 2009), but it produces an inner hole only if the accretion rate falls
below a certain limit and the disk mass is low
enough\footnote{Note that X-ray+FUV photoevaporation, combined with
viscous disk evolution, is efficient removing a large fraction of the disk mass
in a few Myr even if no hole is created.}. The lifetimes of the resulting TO are 
expected to be of the order of few hundred thousand years (Alexander et al. 2006a,b; 
Gorti \& Hollenbach 2009). Other studies involving time-dependent
disk evolution and X-ray/FUV/EUV photoevaporation predict inner hole opening
starting at $\sim$4 Myr, and TO lifetimes that can 
exceed 1 Myr (Gorti et al. 2009), although accretion would drop to undetectable 
levels shortly after the opening of the hole. Photoevaporation also depends strongly
on the structure of the disk and the dust (Gorti \& Hollenbach 2009, Gorti et al. 2009;
Ercolano et al. 2009b). The lower irradiation in flattened
(geometrically thin) disks decreases the efficiency of photoevaporation,
while grain growth decreases the opacity of the disk and thus favors 
photoevaporation. The combined result of these two oposite effects is unclear,
and probably depends on the individual disk parameters (U. Gorti, private 
communication). Some observational studies have suggested that strong 
mass removal by photoevaporation may only be efficient at a late stage in the disk evolution
(Pascucci \& Sterzik 2009), as has been also proposed theoretically (Alexander \& Armitage 2009). 
The theoretical timescales for dust coagulation 
are uncertain (Dullemond \& Dominik 2005; Brauer et al. 2008; Birnstiel et al. 2009),
but observations suggest an important degree of dust evolution in the disks of solar-type
stars by the age of $\sim$4-5 Myr (Sicilia-Aguilar et al. 2006a; Rodmann et al. 2006;
Currie et al. 2009).  

The systematically flatter disks in Tr~37 
(compared to Taurus) and the similar accretion rates of accreting
TO and normal CTTS are consistent with most TO in Tr~37 resulting
from strong grain coagulation that leaves the gas content (and accretion)
relatively unchanged. 
The implications of the lack of statistical differences in the accretion rates
of accreting TO and normal disks in Tr~37 are multiple. On one hand, 
significant amounts of matter must be crossing the depleted inner disks
of accreting TO, which favors the scenario where the near-IR hole
has been created by planetesimal formation/strong grain coagulation, rather 
than being due to gas removal by photoevaporation. A picture where the small
dust has been strongly or even totally removed, but the gas flow
continues more or less undisturbed, seems more complicated to achieve than
a situation where the whole inner disk has been cleared from dust and gas.
Several mechanisms, like dust filtration at the edges of a gap created
by a planet (Rice et al. 2006) and the differences in turbulence/coagulation
between active and dead zones in layered disks (Ciesla 2007) have been proposed
to produce radially-dependent dust distributions and optically thin inner disks.
Radially dependent grain coagulation and settling 
are also consistent with the typical SEDs seen in the CTTS members of Tr~37,
many of which show a change of slope or ``kink'' around 6-8$\mu$m
(Sicilia-Aguilar et al. 2006a). 
Non-accreting TO in Tr~37 cannot be explained by grain coagulation alone,
but require a mechanism to remove the gas or at least, not to allow the
gas transport into the inner disk. Binary companions could be a good candidate, 
although we would expect few surviving circumbinary disks at 4 Myr age (Bouwman et al.
2006), and recent surveys of TO suggest that binaries as the cause of TO are the 
exception and not the rule (Pott et al. 2009). Photoevaporation may only be efficient for the most flared
objects, but strong grain coagulation may help to photoevaporate the gas in the innermost
regions once the dust opacity decreases sufficiently. Finally, the formation
of one or more giant planets may reduce the typically low accretion rates of CTTS
in Tr~37 below our detection limits, which could explain at least part of our
non-accreting TO.

In addition, initial conditions, that have been invoked to explain the
spread in the \.{M} vs. M$_*$ relation (Dullemond et al. 2006), could also determine the
types of disks and TO that dominate in a given cluster. The quiescent, undisturbed
environment in Taurus may affect both the evolution of the
disks, compared to OB associations like Orion L~1630N and 
L~1641 clouds or Tr~37, as well as the initial conditions
at the time of formation (and thus, initial disk masses and sizes).

\subsection{Accretion variability: How common are EX-or objects? \label{variability}}

Young pre-main sequence stars are known to be variable, with most variations
due to the combination of rotation and cool spots (type I variability), 
hot spots and changes in the accretion rate (type II variability), and  obscuration by
circumstellar dust (type III variability; Herbst et al. 1994;
Hillenbrand et al. 1998). In the case of accreting stars, variations up to 
0.5 magnitudes at optical wavelengths in timescales of a few days are common, 
suggesting that the accretion flow is variable (Bertout 1989; 
Brice\~{n}o et al. 2001; Eiroa et al. 2002). Variable accretion has been
also suggested by the changes in the accretion-related emission lines (both in line profile
and in EW; Alencar et al. 2001; Alencar \& Batalha 2002; 
Sicilia-Aguilar et al. 2005a). Accretion variability is more dramatic in
the case of EX-or objects, which experience repeated outbursts of different magnitude
in timescales of months to decades (Hartmann \& Kenyon 1996; Herbig et al. 2001;
Herbig 2007, 2008). Some particularly bright outbursts may
be strong enough to trigger the formation of crystalline dust throughout the
disk (\'{A}brah\'{a}m et al. 2009), which could explain 
the abundance of crystals in the comets in our own Solar System. It has been suggested 
that the EX-or phase should be common for young stars (Hartmann \& Kenyon 1996; 
Hartmann 1998), but the actual fraction of objects suffering accretion
variability is uncertain, especially because the typical variations may 
be too small ($<$2 mag) to be noticed in more distant clusters and associations.

\begin{figure}
\plottwo{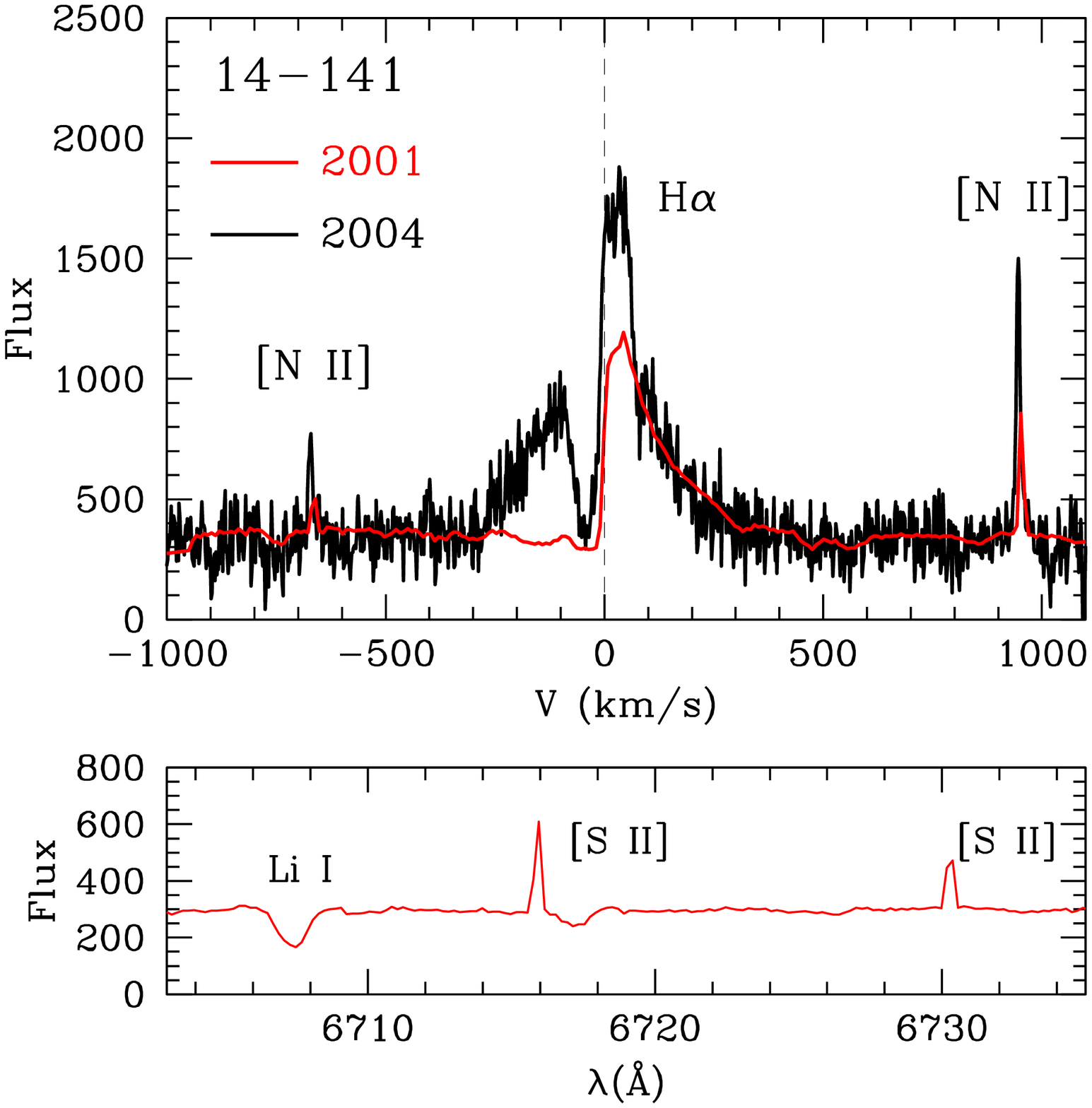}{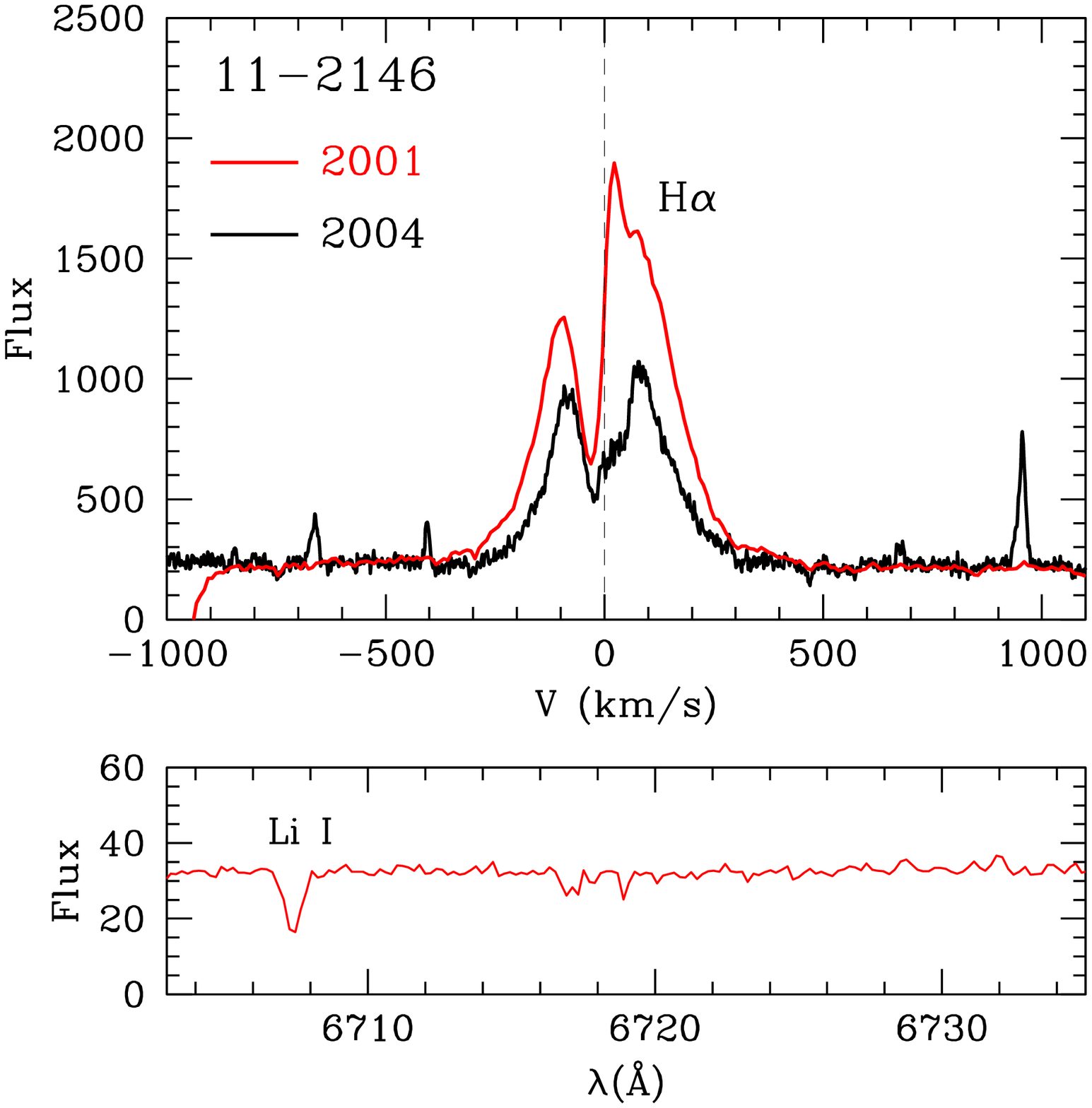}
\caption{H$\alpha$ profile variations of the stars 14-141 and 11-2146. 
The fluxes are given in arbitrary units, scaled to the continuum levels. 
Note that the [N II] lines in 11-2146 are due to the undersubtracted nebular emission
in the multifiber spectrograph Hectospec,
and not related to the object.  \label{spectra}}
\end{figure}

We have discussed in detail the variability of the Tr~37 member GM Cep (also known as
13-277 in our member list; Sicilia-Aguilar et al. 2008a). Taking just into account 
our previous optical data (Sicilia-Aguilar et al. 2004, 2005b) and our new photometry,
we find that more than 10\% of the stars display variations in V of at
least 0.5 magnitudes (see Table \ref{data-table}). Given that we compare here 
only two epochs, we expect that many more of the cluster members regularly
display variability with $\Delta$V$\geq$0.5 mag. A lower degree of variability,
$\Delta$V$\geq$0.1 mag is even more common, with
2/3 (65\%) of the objects showing such changes.   
Among the 8 objects with double measurements in our LAICA survey, we do not
observe significant variations between the \.{M} measured on consecutive days.
Despite the limitations of our FLWO observations, we have compared our 
previous accretion rates with the new values, in order to look for accretion variability 
and optical variability in general. Several objects display significant variations 
in their accretion rates, which can be confirmed in two cases by optical spectroscopy:

The CTTS 11-2146 shows now an accretion rate 
\.{M}$\sim$~10$^{-8}$~M$_\odot$~yr$^{-1}$, 
more than one order of magnitude below the one measured previously 
(\.{M}$\sim$1.7$\times$10$^{-7}$~M$_\odot$~yr$^{-1}$). The variability of the object,
and maybe the presence of a nearby (2") accreting star, could have affected our
less sensitive U band measurements at the FLWO, but available high-resolution
spectra taken with HIRES in 2001 (for details about the HIRES data, see
Sicilia-Aguilar et al. 2008a) and Hectochelle in 2004 (Sicilia-Aguilar et al. 2006b)
are in agreement with \.{M} variations (Figure \ref{spectra}). 
The spectroscopic data reveals changes in the emission line profile as well as 
in the blueshifted absorption, associated to variable accretion plus wind.

The star 14-141 appears to have suffered an increase in its accretion 
rate, compared to our previous observations, where it was not detected. 
The comparison of HIRES 2001 data with Hectochelle spectra taken in 2004 
confirms a very strong H$\alpha$ variability, with remarkable changes
in the line profile (Figure \ref{spectra}). The H$\alpha$ profile in 2001 was dominated by
a shock, being that of a Herbig Haro object (also confirmed by the presence
of forbidden [N II] and [S II] emission; Hartigan et al. 1987). By 2004, the signs of
shock are much weaker, and the profile appears to be typical for an
accreting CTTS, with a blueshifted absorption indicative of a 
wind/outflow.

Although we do not have multiepoch high-resolution spectroscopy for the rest of
the objects, several other stars are candidates for strong optical variability
maybe related to accretion variations: 13-1238, 13-236, 13-1250,
53-1762, 21-1762, 11-2031, and 82-272 (see Table \ref{exors-table}). These
objects are good candidates to look for EX-ors and explore the typical variability
in accreting solar-type stars. In addition, we find a TO with narrow H$\alpha$ in 2004,
and evidence of accretion 
(0.11$^{+0.23}_{-0.06}\times$10$^{-8}$~M$_\odot$~yr$^{-1}$) in 2007, 72-875. 
The star is probably a single-lined spectroscopic
binary, so episodic accretion may be a possibility. Although a few TO
display accretion-related variability (13-1250, 21-1762, 72-875),
we must stress that the most remarkable cases of \.{M} variability correspond
to stars with the most flared and most massive disks (13-277) and objects
associated to the younger ($\sim$1~Myr) part of the cluster (11-2146, 14-141).

Finally, we must mention that several objects are consistent with type III variability,
displaying changes in the extinction larger than expected from the typical errors. While
U and VRI were not simultaneous in our FLWO survey, the three bands VRI were taken 
within 30 min to 1 hour difference, so the A$_V$ estimates are as robust as the 
A$_V$ derived from LAICA observations. Comparing the A$_V$ values in Sicilia-Aguilar et al.
(2004, 2005b) with the LAICA ones, we find several type III variability candidates: 
14-287, 11-581, 13-1048, 11-1067, 82-272, 12-1955, 12-595, 21-840, 21-230, 21-1692,
21-763, 21-1762, and 11-2146 (see Table \ref{uxors-table}).
Among these, 14-287, 13-1048, 82-272, 12-595,  21-840, and 11-2146
have circumstellar disks that could be responsible for the extinction
variations. Therefore, the variability of 11-2146 could be a combination
of type II and type III variability. The stars 13-1048 (M0), 82-272 (G9, SB2
with two similar-mass accreting components),
and 12-595 (K7) have such dramatic A$_V$ changes that might be considered as
good candidates for UX-or variables. In the case of 12-595, the anomalous optical 
colors (with a too blue V-I color for a K7 star, responsible for the 
anomalous age derived from the isochrones) are consistent with a nearly edge-on disk
where scattering is an important contributor to the measured luminosity.

\section{Summary and conclusions \label{conclu}}

We present new optical (UVRI) photometry for the Tr~37 and NGC~7160 clusters
in the Cep~OB2 region. The data allow us to constrain the presence of accretion
and accretion rates in more than 60 T Tauri stars with evolved disks, including
20 TO with zero or very small near-IR excess and presumably, inner holes and/or
coagulated/settled inner disks. 
We study the dependence of the accretion rate on the disk structure (using the
SED slope, $\alpha$, as an indicator of the structure of the disk) finding no 
strong correlation between $\alpha$ at different wavelengths
and the accretion rate. In fact, although we do not find any TO
with a large ($>$10$^{-8}~$M$_\odot$~yr$^{-1}$) accretion rate and
the most flared disks have typically the largest accretion rates, 
strong accretors are rare in regions as old as Tr~37, and we do not 
see any statistically significant difference between the accretion rates
of TO and normal CTTS disks. This result is in contrast with the observations of
Taurus (Najita et al. 2007), which reported that accretion rates for TO
are systematically biased towards lower values of \.{M}, with accretion rates
about one order of magnitude lower than the typical CTTS accretion rates
for a similar disk mass. We must
stress that the accretion rates of TO in Taurus are similar to those of
TO in Cep~OB2, with the difference being that CTTS in Tr~37 have typical
accretion rates about one order of magnitude lower than Taurus CTTS. 
Nevertheless, about half of the TO do not show any evidence of accretion, 
which suggests that the TO class includes several different types of objects and
physical processes. 

The formation of Jovian planets is expected to cause inner holes and reduce the
accretion rate onto the star compared to normal CTTS disks
(Lubow \& D'Angelo 2006; Najita et al. 2007).
On the other hand, photoevaporation by X-ray, EUV, and FUV photons predicts
in general a smaller fraction of TO (especially, a much lower fraction of
accreting TO) and the opening of a hole would happen only when the accretion 
rate drops below a certain limit (10$^{-9}$-10$^{-10}~$M$_\odot$~yr$^{-1}$; 
Alexander et al. 2006b), which also differs from our observations. Nevertheless, 
the efficiency of photoevaporation may be reduced if the disk flaring is low, 
as is the general case in Cep~OB2. New models suggest larger lifetimes for 
photoevaporated TO ($\gtrsim$1 Myr; Gorti et al. 2009), but predict a 
short lifetime of accretion once the hole has been opened. Other 
physical processes, like strong coagulation of the dust, may also result in
optically thin inner disks with relatively normal accretion rates. 

Given the different timescales for the onset of the various disk
clearing mechanisms and the expected lifetimes of each type of TO, it is 
not evident that the TO in all regions are physically the same type 
of objects. The fact that physically different types of TO may dominate at
different ages has been recently suggested from a theoretical point of view
(Alexander \& Armitage 2009). Their models suggest that accreting TO are expected
to contain planets, while most non-accreting TO should have been produced
by photoevaporation. They also claim at at young ages ($<$2 Myr, similar
to Taurus), TO will be dominated by planet formation, while at older stages
($>$6 Myr), photoevaporation should be the main responsible for TO. 
Here we find that in an intermediate-aged cluster like Tr~37, the similarity
of the accretion rates of CTTS and TO are consistent with most of the 
accreting TO being related to strong grain coagulation and settling. In the case of
the TO that do not show any evidence of accretion, other mechanisms,
like the presence of companions (stellar/substellar), giant planet formation,
and/or photoevaporation, may be responsible for their inner holes.
The fact that the fraction of non-accreting TO in Tr~37 is higher($\sim$50\%) 
than in younger regions like Orion ($\sim$30\%; Fang et al. 2009) would also
be in agreement with photoevaporation becoming more efficient at older
ages (Alexander \& Armitage 2009).

Finally, we explore stellar variability with our very limited two-epoch
UVRI observations, finding that $\sim$10\% of the stars show
magnitude variations with $\Delta$V$\geq$0.5 mag, and $\sim$65\% have
$\Delta$V$\geq$0.1 mag. Several cases are compatible with type II variability
(or increased accretion episodes, as in EX-or objects), and many others
show important variations in their extinction, suggesting type III or UX-or-like
variations, sometimes added to a variable accretion. The relatively large number
of type II variables found with data from only two epochs suggests that
accretion variations are very common among young stars. Although the most 
remarkable accretion variations occur in the younger stars with more massive
disks, we also observe accretion variability in some TO. Large variations in
the accretion rate could be an important mechanism to produce crystalline
silicates as those found in the Solar System (\'{A}brah\'{a}m et al. 2009).

Acknowledgments: We want to thank U. Gorti for the insight and discussion about
photoevaporation, especially for the cases of evolved disks, and M. Fang for providing
the accretion data in the Orion suburbs. We are also indebted to C. Mordasini
and H. Klahr for the discussion about the evolution
of disks after planet formation, and the anonymous referee for a detailed
review of our paper that helped clarifying this work. We are also
grateful to B. Braunecker and F. Hormuth
for their help at the 70 cm K\"{o}nigstuhl telescope, as well as the service 
mode observers in Calar Alto, S. Pedraz and J. Aceituno. We  finally thank L. Hillenbrand
for providing the HIRES spectra of the objects 11-2146 and 14-141. 
A.S-A. acknowledges support from the Deutsche Forschungsgemeinschaft, DFG,
grant number SI 1486/1-1.
Based on observations collected at the Centro Astron\'{o}mico Hispano Alem\'{a}n (CAHA) 
at Calar Alto, operated jointly by the Max-Planck Institut f\"{u}r Astronomie and the 
Instituto de Astrof\'{\i}sica de Andaluc\'{\i}a (CSIC).
This work makes use of data products from the Two Micron All Sky Survey, which is 
a joint project of the University of Massachusetts and the Infrared Processing and Analysis 
Center/California Institute of Technology, funded by the National Aeronautics and Space 
Administration and the National Science Foundation.

\clearpage

\clearpage

\clearpage

\clearpage

\clearpage

\clearpage

\begin{deluxetable}{lccc}
\tabletypesize{\scriptsize}
\tablenum{1}
\tablecolumns{4} 
\tablewidth{0pc} 
\tablecaption{Observation Summary \label{obs-table}} 
\tablehead{
 \colhead{Date} & \colhead{Instrument/Telescope} & \colhead{Pointing and Filters}  & \colhead{Seeing} } 
\startdata
2007 June 09 & LAICA/3.5~m & Tr~37-1(UVRI) 			& 1.2-1.3"   \\
2007 June 10 & LAICA/3.5~m & Tr~37-2(UVRI), Tr~37-3(U) 		& 0.8-1.0" \\
2007 June 11 & LAICA/3.5~m & Tr~37-3(VRI), Tr~37-4(UVRI) 	& 0.6-0.8" \\
2007 October 5 & 70~cm KING & NGC~7160 (UVRI)			& 2"   \\
\enddata
\tablecomments{Summary of the UVRI observations. Pointing 2 and U band of pointing 3 were taken
within less than 2 hours time difference. VRI data for pointing 3 and pointing
4 were also taken within less than 2 hours difference.}
\end{deluxetable}

\clearpage

\begin{deluxetable}{lccccccccl}
\tabletypesize{\scriptsize}
\rotate
\tablenum{2}
\tablecolumns{10} 
\tablewidth{0pc} 
\tablecaption{Optical Data \label{data-table}} 
\tablehead{
 \colhead{Name} & \colhead{2MASS ID} &\colhead{Sp. Type} &\colhead{Pointing}& \colhead{U}& \colhead{V}& \colhead{R$_J$}& \colhead{I$_J$}& \colhead{A$_V$}& \colhead{Notes}} 
\startdata
{\bf Tr~37} \\
72-489 & 21351481+5721232 & K5.0 & 4c  & 20.826$\pm$0.081 & 18.159$\pm$0.006 & 16.790$\pm$0.022 & 15.497$\pm$0.004 & 1.76$\pm$0.26 &  n(2")$^a$ \\
72-1427 & 21351627+5728222 & M1.0 & 3c,4c  & 20.176$\pm$0.046 & 18.272$\pm$0.007 & 16.618$\pm$0.022 & 15.161$\pm$0.004 & 1.14$\pm$0.16 &  n(4")$^a$ \\
81-541 & 21351745+5748223 & K5.5 & 4b*  & 20.511$\pm$0.062 & 17.605$\pm$0.015 & 16.086$\pm$0.011 & 14.598$\pm$0.005 & 2.25$\pm$0.29 &  \\
72-875 & 21354975+5724041 & M0.5 &  4c & 20.814$\pm$0.075 & 18.245$\pm$0.007 & 16.590$\pm$0.022 & 15.006$\pm$0.003 & 1.51$\pm$0.27 &  SB1:,n(2.5") \\
21362507 & 21362507+5727502 & M0.0 &  4c & 21.833$\pm$0.188 & 20.426$\pm$0.037 & 18.578$\pm$0.025 & --- & 2.28$\pm$0.25 &  var, SB1  \\
14-141 & 21364941+5731220 & K6.0 &  2b & 19.219$\pm$0.083 & 16.036$\pm$0.002 & 14.508$\pm$0.030 & 13.094$\pm$0.003 & 2.04$\pm$0.24 &   \\
11-2146 & 21365767+5727331 & K6.0 & 4c  & 18.684$\pm$0.014 & 16.662$\pm$0.004 & 15.211$\pm$0.022 & 13.999$\pm$0.002 & 1.62$\pm$0.13 & n(2"+2")$^a$  \\
11-1209 & 21365850+5723257 & K6.0 & 4c  & 17.755$\pm$0.010 & 15.340$\pm$0.003 & 14.040$\pm$0.022 & 13.094$\pm$0.002 & 0.92$\pm$0.09 & n(4")  \\
11-1659 & 21370088+5725224 & K5.0 &  4c & 20.155$\pm$0.042 & 17.046$\pm$0.004 & 15.625$\pm$0.022 & 14.534$\pm$0.003 & 1.72$\pm$0.09 &   \\
11-1499 & 21370140+5724458 & M1.5 & 4c  & 21.062$\pm$0.094 & 17.854$\pm$0.005 & 16.185$\pm$0.022 & 14.424$\pm$0.003 & 1.26$\pm$0.40 &  var,n(5")  \\
11-2322 & 21370191+5728222 & M1.0 & 4c  & 18.858$\pm$0.016 & --- & 15.357$\pm$0.022 & 13.953$\pm$0.002 & 1.30$\pm$0.23 &  var,n(2")$^a$  \\
14-222 & 21370607+5732015 & K7.0 &  1a & 18.926$\pm$0.059 & 15.801$\pm$0.004 & 14.494$\pm$0.011 & 13.257$\pm$0.003 & 0.90$\pm$0.28 &   \\
"       & 21370607+5732015 & K7.0 & 2b  & 18.828$\pm$0.082 & 15.847$\pm$0.002 & 14.404$\pm$0.030 & 13.140$\pm$0.003 & 1.33$\pm$0.20 &   \\
14-287 & 21370649+5732316 & M0.0 & 1a  & 18.858$\pm$0.056 & 17.095$\pm$0.003 & 15.780$\pm$0.004 & 14.385$\pm$0.004 & 1.11$\pm$0.25 &  var  \\
"       & 21370649+5732316 & M0.0 & 2b  & 19.118$\pm$0.082 & 17.258$\pm$0.008 & 15.763$\pm$0.030 & 14.361$\pm$0.003 & 1.11$\pm$0.28 &  var  \\
11-2037 & 21370703+5727007 & K4.5 & 3c  & 18.871$\pm$0.010 & --- & --- & 13.687$\pm$0.002 & 1.63$\pm$0.45 &    \\
11-1067 & 21370843+5722484 & M0.5 & 3c  & 21.635$\pm$0.083 & 17.860$\pm$0.017 & 16.361$\pm$0.012 & --- & 0.64$\pm$0.12 &   \\
14-11 & 21371031+5730189 & M1.5 & 4a  & 20.899$\pm$0.259 & 17.653$\pm$0.004 & 16.170$\pm$0.005 & 14.269$\pm$0.003 & 0.85$\pm$0.72 &   \\
14-125 & 21371054+5731124 & K5.0 & 2b  & 19.163$\pm$0.082 & 16.616$\pm$0.003 & 15.224$\pm$0.030 & 14.041$\pm$0.003 & 1.72$\pm$0.16 &   \\
"       & 21371054+5731124 & K5.0 & 3c,2b  & 19.896$\pm$0.019 & 16.616$\pm$0.003 & 15.224$\pm$0.030 & 14.041$\pm$0.003 & 1.72$\pm$0.16 &   \\
11-1513 & 21371183+5724486 & K7.5 & 3c  & 20.988$\pm$0.047 & 17.068$\pm$0.016 & 15.595$\pm$0.012 & 14.191$\pm$0.002 & 1.31$\pm$0.27 &   \\
11-2131 & 21371215+5727262 & K6.5 & 3c  & 20.850$\pm$0.041 & --- & 15.853$\pm$0.012 & 14.605$\pm$0.003 & 1.87$\pm$0.23&  var  \\
11-2031 & 21371591+5726591 & K2.0 & 3c  & 18.328$\pm$0.006 & --- & 14.240$\pm$0.012 & 13.345$\pm$0.002 & 1.53$\pm$0.09 &    \\
14-103 & 21371976+5731043 & K7.0 & 1a  & 21.856$\pm$0.623 & 18.145$\pm$0.005 & 16.769$\pm$0.006 & 15.350$\pm$0.007 & 1.28$\pm$0.38 &   \\
14-197 & 21372368+5731538 & K5.5 &  1a & 19.544$\pm$0.085 & 16.864$\pm$0.003 & 15.583$\pm$0.011 & 14.452$\pm$0.004 & 1.20$\pm$0.19 & SB1:  \\
11-581 & 21372828+5720326 & G9.0 &  3c & 20.508$\pm$0.030 & 16.697$\pm$0.016 & 15.316$\pm$0.012 & 14.159$\pm$0.002 & 3.19$\pm$0.09 &   \\
14-1017 & 21372894+5736042 & M0.0 &  3c,1a & 20.719$\pm$0.039 & 18.399$\pm$0.005 & 16.874$\pm$0.006 & 15.226$\pm$0.006 & 1.44$\pm$0.45 &   \\
"       & 21372894+5736042 & M0.0 & 1a  & 21.051$\pm$0.296 & 18.399$\pm$0.005 & 16.874$\pm$0.006 & 15.226$\pm$0.006 & 1.44$\pm$0.45 &   \\
14-335 & 21372915+5732534 & K6.5 &  1a & 21.054$\pm$0.313 & 17.542$\pm$0.004 & 16.100$\pm$0.005 & 14.657$\pm$0.005 & 1.66$\pm$0.33 & n(4")  \\
83-343  & 21373696+5755149 & M0.5 &  2b & 21.040$\pm$0.123 & 17.056$\pm$0.009 & 15.455$\pm$0.008 & 14.077$\pm$0.003 & 1.16$\pm$0.12 & n(5") \\
14-183 & 21373849+5731408 & K6.0 & 1a  & 19.838$\pm$0.105 & 17.112$\pm$0.003 & 15.648$\pm$0.012 & 14.431$\pm$0.004 & 1.67$\pm$0.10 &  n(4"), SB1:  \\
14-2148 & 21374184+5740400 & M1.5 & 1a  & 20.993$\pm$0.304 & 18.260$\pm$0.005 & 16.731$\pm$0.006 & 15.162$\pm$0.006 & 0.67$\pm$0.35 &   \\
11-1384 & 21374486+5724135 & K6.5 & 3c  & 20.630$\pm$0.035 & 17.061$\pm$0.016 & 15.680$\pm$0.012 & 14.388$\pm$0.002 & 1.33$\pm$0.26 & n(5"+5")  \\
13-924 & 21375018+5733404 & K5.0 & 1a  & 20.582$\pm$0.197 & 17.125$\pm$0.003 & 15.788$\pm$0.004 & 14.472$\pm$0.004 & 1.69$\pm$0.30 &   \\
12-1984 & 21375022+5725487 & K6.0 & 3c  & 20.008$\pm$0.020 & --- & 15.561$\pm$0.012 & 14.372$\pm$0.002 & 0.88$\pm$0.05 &    \\
12-2519 & 21375107+5727502 & K5.5 & 3c  & 21.266$\pm$0.057 & --- & 15.958$\pm$0.012 & 14.763$\pm$0.003 & 1.93$\pm$0.17 &    \\
12-1968 & 21375487+5726424 & K6.0 & 3c  & 20.257$\pm$0.025 & --- & 15.556$\pm$0.012 & 14.396$\pm$0.003 & 1.69$\pm$0.26 &  SB2:  \\
12-1422 & 21375756+5724197 & M0.0 &  3c & 22.215$\pm$0.139 & 18.907$\pm$0.019 & 17.312$\pm$0.013 & 15.621$\pm$0.004 & 1.68$\pm$0.43 &   \\
12-1091 & 21375762+5722476 & K2.0 & 3c  & 19.710$\pm$0.015 & 16.274$\pm$0.016 & 14.984$\pm$0.012 & 13.984$\pm$0.002 & 2.33$\pm$0.12 &   \\
13-269 & 21375812+5731199 & K6.5 & 1a  & 19.666$\pm$0.089 & 16.658$\pm$0.005 & 15.332$\pm$0.004 & 13.988$\pm$0.004 & 1.22$\pm$0.35 &   \\
12-583  & 21375827+5720354 & M0.0 &  3c & 21.241$\pm$0.059 & 17.650$\pm$0.016 & 16.093$\pm$0.012 & 14.472$\pm$0.003 & 1.50$\pm$0.40 &  n(5") \\
13-1238 & 21375926+5736162 & M1.0 & 1a  & 19.259$\pm$0.070 & 17.902$\pm$0.004 & 16.337$\pm$0.006 & 14.601$\pm$0.004 & 1.15$\pm$0.49 &   \\
12-2373 & 21380058+5728253 & M1.0 & 3c  & 21.776$\pm$0.092 & --- & 16.214$\pm$0.012 & 14.759$\pm$0.003 & 1.49$\pm$0.18 &  var  \\
82-272 & 21380350+5741349 & G9.0 & 1a  & 19.414$\pm$0.080 & 16.616$\pm$0.005 & 15.266$\pm$0.011 & 14.019$\pm$0.004 & 3.01$\pm$0.06 &  var, SB2  \\
12-1081 & 21380593+5722438 & M0.5 &  3c & 21.989$\pm$0.114 & 18.308$\pm$0.017 & 16.836$\pm$0.012 & 15.438$\pm$0.004 & 0.79$\pm$0.28 &   \\
13-1161 & 21380772+5735532 & M0.0 &  1a & 20.845$\pm$0.241 & 17.974$\pm$0.004 & 16.515$\pm$0.006 & 15.091$\pm$0.004 & 1.02$\pm$0.31 &   \\
12-1613 & 21380848+5725118 & M1.0 &  3c & 22.129$\pm$0.123 & 18.326$\pm$0.017 & 16.727$\pm$0.012 & 15.185$\pm$0.004 & 1.06$\pm$0.28 &   \\
13-1426 & 21380856+5737076 & M0.0 & 1a  & 19.858$\pm$0.103 & 19.174$\pm$0.009 & 17.599$\pm$0.013 & 15.748$\pm$0.005 & 1.78$\pm$0.60 &   \\
13-669 & 21380928+5733262 & K1.0 & 1a  & 18.231$\pm$0.046 & 15.834$\pm$0.004 & 14.630$\pm$0.011 & 13.507$\pm$0.004 & 2.17$\pm$0.12 &   \\
13-350 & 21381384+5731414 & M1.0 & 1a  & 21.078$\pm$0.304 & 18.463$\pm$0.006 & 16.880$\pm$0.014 & 15.470$\pm$0.005 & 0.89$\pm$0.17 &  n(3") \\
54-1781 & 21381612+5719357 & M1.0 & 3c  & 21.026$\pm$0.045 & 18.564$\pm$0.018 & 16.922$\pm$0.012 & 15.120$\pm$0.003 & 1.44$\pm$0.47 &   \\
13-1877 & 21381703+5739265 & K7.0 & 1a  & 18.883$\pm$0.057 & 17.189$\pm$0.004 & 15.620$\pm$0.004 & 14.222$\pm$0.004 & 1.84$\pm$0.16 &   \\
13-277 & 21381731+5731220 & G8.0 & 1a  & 15.016$\pm$0.040 & 13.146$\pm$0.004 & 12.090$\pm$0.011 & 11.117$\pm$0.003 & 1.95$\pm$0.09 &  GM Cep, var, SB1:  \\
12-1009 & 21381750+5722308 & K5.5 & 3c  & 20.088$\pm$0.023 & 16.689$\pm$0.016 & 15.373$\pm$0.012 & 14.235$\pm$0.003 & 1.31$\pm$0.18 &  var,n(4"+4")  \\
13-819 & 21382596+5734093 & K5.5 & 1a  & 19.128$\pm$0.065 & 16.423$\pm$0.003 & 15.173$\pm$0.004 & 14.008$\pm$0.004 & 1.14$\pm$0.25 &   \\
12-1955 & 21382692+5726385 & K6.5 &  3c & 21.015$\pm$0.046 & 17.388$\pm$0.009 & 16.020$\pm$0.012 & --- & 1.04$\pm$0.09 &   \\
13-236 & 21382742+5731081 & K2.0 & 1a  & 17.665$\pm$0.042 & 15.546$\pm$0.004 & 14.431$\pm$0.011 & 13.412$\pm$0.004 & 1.63$\pm$0.13 &   \\
13-157 & 21382804+5730464 & K5.5 & 3c  & 18.588$\pm$0.008 & 16.768$\pm$0.016 & 15.320$\pm$0.012 & 14.012$\pm$0.004 & 1.86$\pm$0.20 &  var,n(4")  \\
13-232 & 21382834+5731072 & M0.0 & 1a  & 21.118$\pm$0.310 & 17.428$\pm$0.004 & 16.087$\pm$0.006 & 14.657$\pm$0.004 & 0.68$\pm$0.44 &   \\
21383216 & 21383216+5726359 & M0.0 & 3c  & 21.598$\pm$0.076 & 19.911$\pm$0.015 & 18.202$\pm$0.016 & 16.287$\pm$0.007 & 2.24$\pm$0.52 &    \\
13-566 & 21383481+5732500 & K5.5 & 1a  & 20.977$\pm$0.279 & 18.030$\pm$0.004 & 16.555$\pm$0.006 & 14.959$\pm$0.005 & 2.22$\pm$0.43 &   \\
13-1709 & 21384038+5738374 & K5.5 & 1a  & 20.832$\pm$0.251 & 16.878$\pm$0.003 & 15.591$\pm$0.011 & 14.381$\pm$0.003 & 1.29$\pm$0.26 &   \\
54-1613 & 21384332+5718359 & K5.0 & 3c  & 20.014$\pm$0.020 & --- & 15.195$\pm$0.012 & 14.277$\pm$0.002 & 1.06$\pm$0.09 &    \\
21384350 & 21384350+5727270 & M2.0 & 3c  & 24.093$\pm$0.775 & 19.918$\pm$0.013 & --- & --- & 1.67$\pm$0.45 &  var  \\
54-1547 & 21384446+5718091 & K5.5 & 3c  & 19.168$\pm$0.010 & 16.571$\pm$0.007 & 15.321$\pm$0.012 & 14.396$\pm$0.003 & 0.91$\pm$0.06 &   \\
"       & 21384446+5718091 & K5.5 & 4d  & 18.539$\pm$0.027 & 16.571$\pm$0.007 & 15.299$\pm$0.011 & 14.437$\pm$0.007 & 0.91$\pm$0.08 &   \\
12-2363 & 21384544+5728230 & M0.5 & 3c  & 21.420$\pm$0.077 & 17.552$\pm$0.009 & --- & --- & 1.88$\pm$0.45 & n(2"), SB1:  \\
12-595  & 21384622+5720380 & K7.0 & 3c  & 22.047$\pm$0.118 & 19.512$\pm$0.011 & 17.497$\pm$0.013 & 16.144$\pm$0.006 & 3.12$\pm$0.36 &  var,n(4")  \\
12-1423 & 21384707+5724207 & K7.0 &  3c & 20.814$\pm$0.039 & 17.042$\pm$0.016 & 15.601$\pm$0.012 & 14.382$\pm$0.002 & 1.28$\pm$0.14 &   \\
12-1010  & 21385029+5722283 & M2.0 & 3c  & 20.896$\pm$0.039 & 18.554$\pm$0.009 & 17.043$\pm$0.012 & 15.463$\pm$0.004 & 0.41$\pm$0.37 &   \\
12-2098 & 21385253+5727184 & M2.5 & 3c  & 22.178$\pm$0.152 & 18.775$\pm$0.010 & --- & --- & 0.98$\pm$0.45 &   \\
91-506 & 21385807+5743343 & K6.5 &  4a* & 19.777$\pm$0.063 & 17.141$\pm$0.013 &  15.732$\pm$0.017 &  14.546$\pm$0.004  & 1.32$\pm$0.15 & \\
12-1617 & 21390468+5725128 & M1.0 & 4d  & 20.831$\pm$0.089 & 18.039$\pm$0.008 & --- & --- & 1.60$\pm$0.45 &   \\
21-840 & 21391012+5722323 & M1.0 & 4d  & 21.201$\pm$0.127 & 18.852$\pm$0.011 & --- & 15.647$\pm$0.008 & 1.45$\pm$0.04 &   \\
13-1048 & 21391088+5735181 & M0.0 & 2a  & 19.754$\pm$0.039 & 17.081$\pm$0.012 & --- & 14.523$\pm$0.003 & 0.71$\pm$0.03 &  var  \\
13-1250 & 21391213+5736164 & K4.5 & 2a  & 18.403$\pm$0.018 & 15.956$\pm$0.012 & --- & 13.668$\pm$0.003 & 1.42$\pm$0.03 &   \\
21-563 & 21391288+5721088 & M1.0 &  4d & 20.807$\pm$0.085 & 17.501$\pm$0.008 & --- & --- & 1.25$\pm$0.45 &   \\
24-542 & 21392957+5733417 & K4.0 & 2a  & 18.548$\pm$0.018 & 15.986$\pm$0.012 & 14.675$\pm$0.032 & 13.718$\pm$0.003 & 1.53$\pm$0.14 &   \\
24-515 & 21393407+5733316 & M0.5 & 2a  & 19.908$\pm$0.043 & 17.941$\pm$0.012 & 16.396$\pm$0.032 & 15.123$\pm$0.004 & 0.89$\pm$0.15 &   \\
21-998 & 21393480+5723277 & K5.5 &  4d & 19.798$\pm$0.040 & 17.668$\pm$0.008 & --- & 14.961$\pm$0.007 & 1.95$\pm$0.03 &  SB1: \\
21-33 & 21393561+5718220 & M0.0 &  4d & 20.814$\pm$0.075 & 18.723$\pm$0.011 & 17.138$\pm$0.011 & 15.722$\pm$0.008 & 1.39$\pm$0.18 &   \\
24-48 & 21393805+5730439 & M0.5 &  2a & 21.972$\pm$0.279 & 17.996$\pm$0.012 & --- & --- & 1.50$\pm$0.45 &   \\
21-230 & 21394169+5719274 & M0.5 &  4d & 21.044$\pm$0.103 & 17.922$\pm$0.008 & --- & 15.234$\pm$0.007 & 0.71$\pm$0.03 &   \\
21-1536 & 21394570+5726242 & M0.0 & 4d  & 20.636$\pm$0.078 & 19.048$\pm$0.013 & --- & --- & 1.77$\pm$0.45 &   \\
52-1649 & 21394643+5705072 & K5.0 & 2d*  & 20.051$\pm$0.035 & 17.264$\pm$0.007 & 15.985$\pm$0.007 & 15.043$\pm$0.006 &  1.16$\pm$0.04 & \\
21-2251 & 21394754+5725210 & M2.0 &  4d & 21.175$\pm$0.115 & 18.020$\pm$0.008 & --- & --- & 1.50$\pm$0.45 &  SB1 \\
21-1586 & 21394793+5726427 & K7.0 &  4d & 21.509$\pm$0.156 & 18.402$\pm$0.009 & --- & --- & 1.49$\pm$0.45 &   \\
53-1762 & 21395029+5719177 & M0.0 & 4d  & 21.391$\pm$0.141 & 18.453$\pm$0.009 & --- & 15.655$\pm$0.008 & 1.17$\pm$0.03 &  var, SB1:  \\
21-1692 & 21400128+5727184 & M1.0 & 4d  & 21.645$\pm$0.170 & 18.239$\pm$0.009 & --- & 15.386$\pm$0.007 & 0.77$\pm$0.03 &   \\
21-763 & 21400259+5722090 & M0.0 & 4d  & 20.511$\pm$0.067 & 17.456$\pm$0.008 & --- & 14.950$\pm$0.007 & 0.61$\pm$0.03 &  n(5") \\
21-895a & 21400321+5722505 & K5.0 & 4d  & 18.983$\pm$0.028$^b$ & 16.640$\pm$0.007 & --- & 14.305$\pm$0.007 & 1.37$\pm$0.03 &  n(2")$^b$ \\
21400451 & 21400451+5728363 & K5.0 & 2a  & 18.045$\pm$0.018 & 16.165$\pm$0.012 & --- & 13.893$\pm$0.003 & 1.25$\pm$0.03 &  n(5"+5")  \\
"       & 21400451+5728363 & K5.0 &  4d & 18.215$\pm$0.024 & 16.304$\pm$0.007 & --- & 14.088$\pm$0.007 & 1.14$\pm$0.03 & n(5"+5")  \\
21-1762 & 21400924+5727393 & K5.0 & 4d  & 19.753$\pm$0.039 & 16.955$\pm$0.007 & --- & 14.519$\pm$0.007 & 1.57$\pm$0.03 &   \\
24-1736 & 21401134+5739518 & M1.0 & 2a  & 19.886$\pm$0.045 & 18.762$\pm$0.013 & 17.138$\pm$0.032 & 15.497$\pm$0.004 & 1.23$\pm$0.36 &   \\
24-1796 & 21401182+5740121 & K7.0 & 2a  & 18.389$\pm$0.018 & 17.341$\pm$0.012 & 15.897$\pm$0.032 & 14.825$\pm$0.004 & 1.15$\pm$0.14 & SB2:  \\
21-2006 & 21401390+5728481 & K5.0 &  2a & 19.564$\pm$0.034 & 16.933$\pm$0.012 & 15.517$\pm$0.032 & 14.589$\pm$0.004 & 1.55$\pm$0.21 &   \\
"       & 21401390+5728481 & K5.0 & 4d  & 19.457$\pm$0.033 & 16.926$\pm$0.007 & 15.517$\pm$0.032 & 14.707$\pm$0.007 & 1.42$\pm$0.30 &   \\
22-2651 & 21402130+5726579 & M1.5 & 4d  & 18.825$\pm$0.026 & 17.878$\pm$0.008 & --- & 15.375$\pm$0.007 & 1.67$\pm$0.45$^c$ &  var  \\
22-1418 & 21402287+5727329 & M1.5 & 4d  & 20.799$\pm$0.082 & 17.861$\pm$0.008 & --- & --- & 0.73$\pm$0.45 &   \\
23-405 & 21403134+5733417 & K5.0 & 2a  & 19.142$\pm$0.025 & 16.610$\pm$0.012 & 15.209$\pm$0.032 & 14.271$\pm$0.003 & 1.52$\pm$0.19 &   \\
23-570 & 21403574+5734550 & K6.0 & 2a  & 18.391$\pm$0.017 & 16.718$\pm$0.012 & 15.276$\pm$0.032 & 14.308$\pm$0.003 & 1.36$\pm$0.19 &   \\
\\
{\bf NGC~7160}\\
01-580 & 21533707+6228469 & K4.5 & ---  & 18.38$\pm$0.15 & 17.392$\pm$0.042 & 15.755$\pm$0.020 & 14.737$\pm$0.015 & 2.46$\pm$0.40 &  \\
\enddata
\tablecomments{Optical data and 2MASS IDs for the studied stars. 
Due to the small field overlap, some stars were observed more than once. 
The pointing and LAICA CCD ID (a,b,c,d) are indicated to show whether all the UVRI measurements
were simultaneous (see Table \ref{obs-table}). Two pointings are indicated (e.g. 3c+2b),
when the U magnitude was observed with pointing 3c, and the VRI data
comes from pointing 2. 
Notes: "var" = variability ($\Delta$V$>$0.5 mags) observed, 
"n(x")" = nearby star at x" projected distance, usually the photometry 
is not affected. If the nearby star is bright and may affect the photometry,
$^a$ is added (e.g., n(2")$^a$). Objects that appeared merged with other stars/artifacts
are excluded from this study. $^b$ 21-895b is probably not a member and not a physical companion 
of 21-895a because of its position in the color-magnitude diagram. $^c$ The colors of
22-2651 are probably affected by non-standard variability/anomalous extinction. 
Therefore, we adopt for this object the cluster average extinction. }
\end{deluxetable}

\clearpage

\begin{deluxetable}{lcccccccccc}
\tabletypesize{\scriptsize}
\rotate
\tablenum{3}
\tablecolumns{9} 
\tablewidth{0pc} 
\tablecaption{Accretion and Stellar Properties\label{acc-table}} 
\tablehead{
 \colhead{Name} & \colhead{H$\alpha$ EW(\AA)} & \colhead{H$\alpha$ profile} &\colhead{Disk}& \colhead{L (L$_\odot$)} & \colhead{R (R$_\odot$)}& \colhead{Age (Myr)}& \colhead{M (M$_\odot$)}& \colhead{\.{M}$_U$ }  &  \colhead{\.{M}$_{fin}$}  &  \colhead{Comments} }
\startdata
{\bf Tr~37} \\
72-489 & (-5.5) & --- & --- & 0.38 & 1.1 & 22.9 & 0.9 & 0.04$^{+0.03}_{-0.01}$ & 0.04 &  \\   
72-1427 & -16,(-77) & B & TO & 0.36 & 1.5 & 3.3 & 0.5 & 0.15$^{+0.28}_{-0.07}$ & 0.15 &  \\   
81-541 & -31,(-60) & B & CTTS & 1.07 & 1.9 & 3.6 & 1.0 & 0.19$^{+0.15}_{-0.06}$ & 0.19 & \\   
72-875 & (-21) & N & TO & 0.50 & 1.7 & 2.6 & 0.5 & 0.11$^{+0.23}_{-0.06}$ & 0.11 & SB1:, now accreting \\ 
21362507 & -78,(-86) & B & CTTS & 0.11 & 0.8 & 26.0 & 0.6 & 0.09$^{+0.16}_{-0.05}$ & 0.09 & SB1, HR \\  
14-141 & -15,(-5) & B & CTTS & 3.79 & 3.7 & 0.8 & 0.9 & 0.62$^{+0.57}_{-0.33}$ & $<$2.3 & \\    
11-2146 & -28,-86,(-33) & B & CTTS & 1.37 & 2.3 & 2.5 & 1.0 & 0.99$^{+0.67}_{-0.37}$ & 0.99 &  \\  
11-1209 & -6,(-4) & B & CTTS & 2.31 & 2.9 & 1.4 & 1.0 &  0.56$^{+0.35}_{-0.23}$  & 0.56 &\\    
11-1659 & (-2) & N & N & 0.91 & 1.7 & 8.3 & 1.1 & 0.02$^{+0.02}_{-0.02}$ & 0 &   \\  
11-1499 & (-7) & N & N & 0.73 & 2.2 & 1.4 & 0.4 & 0.00$^{+0.12}_{-0.02}$ & 0 &  \\  
11-2322 & -18,(-23) & B & CTTS & 1.37 & 2.9 & 0.7 & 0.5 & 1.6$^{+2.9}_{-0.8}$ & 1.6 &  \\   
14-222 & (-5) & N: & TOe8 & 1.87 & 2.8 & 1.1 & 0.7 & -0.06$^{+0.03}_{-0.08}$ & $<$0.04 &   \\  
" & (-5) & N: & TOe8 & 2.51 & 3.3 & 0.9 & 0.8 & 0.34$^{+0.28}_{-0.25}$ & $<$0.85 &   \\  
14-287 & -18,(-35) & B & CTTS & 0.61 & 1.8 & 2.3 & 0.5 & 0.21$^{+0.34}_{-0.11}$ & 0.21 &  \\   
" & -18,(-35) & B & CTTS & 0.79 & 2.0 & 2.0 & 0.6 & 0.46$^{+0.73}_{-0.24}$ & 0.46 & \\   
11-2037 & -43,(-50) & B & CTTS & 1.88 & 2.3 & 2.5 & 1.3 & 1.1$^{+2.6}_{-0.8}$ & $<$9 & HR \\  
11-1067 & (-7) & N & N & 0.28 & 1.3 & 5.4 & 0.5 & 0.00$^{+0.02}_{-0.01}$ & 0 &   HR \\ 
14-11 & (-5) & N & TO & 0.70 & 2.1 & 1.4 & 0.4 & -0.05$^{+0.07}_{-0.03}$ & 0 &   \\  
14-125 & -14,(-13) & B & CTTS & 1.44 & 2.2 & 3.0 & 1.1 & 0.34$^{+0.26}_{-0.12}$ & 0.34 &  \\   
" & -14,(-13) & B & CTTS & 1.44 & 2.2 & 3.0 & 1.1 & 0.34$^{+0.21}_{-0.12}$ & 0.34 &  \\    
11-1513 & (-5) & N & N & 0.97 & 2.1 & 2.0 & 0.6 & 0.02$^{+0.06}_{-0.03}$ & 0 &   \\  
11-2131 & (-10) & B & CTTS & 0.78 & 1.8 & 3.0 & 0.8 & 0.25$^{+0.12}_{-0.12}$ & 0.25 &  HR \\  
11-2031 & -5(-5) & B & CTTS & 2.93 & 2.4 & 4.5 & 1.6 & 0.27$^{+0.23}_{-0.17}$ & $<$1 &  HR \\  
14-103 & (-2) & N & N & 0.32 & 1.2 & 11.6 & 0.8  & -0.02$^{+0.03}_{-0.02}$ & 0 &  \\    
14-197 & (-2) & N & TOe8 & 0.77 & 1.6 & 7.2 & 1.0 & 0.03$^{+0.04}_{-0.01}$ & 0/$<$0.14 & SB1 \\  
11-581 & (-9) & N & N & 3.39 & 2.1 & 24.2 & 1.4 & -0.51$^{+0.36}_{-0.50}$ & 0 &  G-age\\   
14-1017 & (-55) & B & CTTS & 0.41 & 1.5 & 3.5 & 0.5 & 0.29$^{+0.41}_{-0.14}$ & $<$1.5 &  \\    
" & (-55)& B & CTTS & 0.41 & 1.5 & 3.5 & 0.5 & 0.05$^{+0.14}_{-0.03}$ & $<$0.5 &   \\   
14-335 & (-20) & P & CTTS & 0.74 & 1.7 & 4.7 & 0.8 & -0.02$^{+0.05}_{-0.05}$ & $<$0.14 &  \\  
83-343 & (-3) & P & --- & 1.01 & 2.4 & 1.2 & 0.5 & -0.12$^{+0.03}_{-0.05}$ & $<$0.1 &  \\  
14-183 & -14,(-65) & B & CTTS & 0.94 & 1.9 & 4.8 & 1.0 & 0.16$^{+0.12}_{-0.07}$ & 0.16 &  \\   
14-2148 & (-2) & N & TOe8 & 0.29 & 1.4 & 3.4 & 0.4 & 0.01$^{+0.05}_{-0.01}$ & 0 &    \\  
11-1384 & (-5) & N & N & 0.82 & 1.8 & 4.3 & 0.9 & 0.06$^{+0.05}_{-0.04}$ & 0 &   \\   
13-924 & (-4) & N & N & 0.95 & 1.8 & 5.5 & 1.0 & -0.08$^{+0.04}_{-0.09}$ & 0 &   \\   
12-1984 & (-5) & N & N & 0.99 & 1.9 & 2.7 & 0.9 & 0.64$^{+0.31}_{-0.23}$ & 0/0.6 &   HR \\  
12-2519 & (-8) & P & CTTS & 0.71 & 1.6 & 2.8 & 0.9 & 0.07$^{+0.03}_{-0.03}$ & 0.07 &   HR \\  
12-1968 & -8,(-11) & B & CTTS & 0.97 & 1.9 & 2.8 & 0.9 & 0.34$^{+0.15}_{-0.13}$ & 0.34 &  SB2, HR \\  
12-1422 & (-17) & N & N:$^*$ & 0.32 & 1.3 & 5.8 & 0.5 & 0.05$^{+0.10}_{-0.03}$ & $<$0.24 &  large hole? \\   
12-1091 & -2,(-17) & B & CTTS & 2.64 & 2.3 & 14.1 & 1.4 & 0.08$^{+0.12}_{-0.07}$ & $<$0.4 &   \\  
13-269 & (-7) & P & N & 1.12 & 2.1 & 2.4 & 0.8 & 0.01$^{+0.04}_{-0.03}$ & 0/$<$0.1 &   \\  
12-583 & (-7) & N & N & 0.85 & 2.1 & 1.6 & 0.5 & 0.11$^{+0.25}_{-0.07}$ & 0 &   \\   
13-1238 & -31,(-64) & B & CTTS & 0.60 & 1.9 & 1.8 & 0.4 & 0.6$^{+1.1}_{-0.3}$ & 0.6 &   \\   
12-2373 & (-6) & N & N & 0.65 & 2.0 & 1.6 & 0.5 & 0.07$^{+0.21}_{-0.05}$ & 0 &   HR \\ 
82-272 & -15,(-13) & B & CTTS & 3.03 & 2.0 & 12.7 & 1.5 & 0.07$^{+0.27}_{-0.34}$ & $<$0.9 &   SB2, G-age \\ 
12-1081 & (-4) & N & TO & 0.24 & 1.2 & 7.2 & 0.5 & 0.00$^{+0.02}_{-0.01}$ & 0 &    \\  
13-1161 & (-1) & N & N & 0.39 & 1.4 & 4.9 & 0.6 & 0.01$^{+0.06}_{-0.01}$ & 0 &    \\  
12-1613 & (-13) & N & N & 0.34 & 1.4 & 3.3 & 0.5 & 0.01$^{+0.04}_{-0.01}$ & 0 &   \\  
13-1426 & -40,(-109) & B & CTTS & 0.30 & 1.2 & 5.6 & 0.5 & 0.6$^{+0.8}_{-0.3}$ & 0.6 &   \\   
13-669 & -18,(-22) & B & CTTS & 3.23 & 2.4 & 6.3 & 1.6 & 0.58$^{+0.20}_{-0.23}$ & 0.58 &   \\   
13-350 & (-9) & N & TO & 0.24 & 1.2 & 6.4 & 0.5 & 0.02$^{+0.06}_{-0.01}$ & 0 &    \\  
54-1781 & (-13) & P & TO & 0.42 & 1.6 & 2.5 & 0.4 & 0.3$^{+0.5}_{-0.1}$ & 0.3 &   \\   
13-1877 & -33,(-68) & B & CTTS & 1.16 & 2.2 & 2.3 & 0.8 & 1.6$^{+1.2}_{-0.8}$ & 1.6 &  \\   
13-277 & -14,-6,(-10,-19) & B & CTTS & 27.75 & 5.9 & 1.9 & 2.6 & 14$^{+6}_{-4}$ & 14 &  GM Cep, SB1: \\   
12-1009 & (-4) & N & TO: & 0.98 & 1.8 & 5.0 & 1.0 & 0.07$^{+0.06}_{-0.02}$ & 0.07 &   \\   
13-819 & -6,(-10) & B & TO:$^*$ & 1.12 & 2.0 & 3.2 & 1.0 & 0.03$^{+0.04}_{-0.02}$ & $<$0.14 & large hole   \\  
12-1955 & (-2) & N & N & 0.41 & 1.3 & 9.2 & 0.9 & 0.01$^{+0.01}_{-0.01}$ & 0 &   \\    
13-236 & -56,(-47) & B & CTTS & 2.70 & 2.3 & 5.4 & 1.6 & 0.51$^{+0.26}_{-0.15}$ & 0.51 &   \\   
13-157 & -20,(-14) & B & CTTS & 1.54 & 2.3 & 2.4 & 1.0 & 4.2$^{+2.9}_{-1.5}$ & 4.2 &   \\   
13-232 & (-3) & N & N: & 0.50 & 1.6 & 2.7 & 0.5 & -0.05$^{+0.02}_{-0.03}$ & 0 &   \\   
21383216 & (-129) & B & CTTS & 0.22 & 1.1 & 8.4 & 0.5 & 0.4$^{+0.5}_{-0.2}$ & 0.4 &  \\   
13-566 & (-5) & N & TOe8 & 0.75 & 1.6 & 6.3 & 0.9 & 0.08$^{+0.14}_{-0.06}$ & 0 &   \\  
13-1709 & (-3) & N & N & 0.85 & 1.7 & 5.8 & 1.0 & -0.14$^{+0.06}_{-0.10}$ & 0 &   \\   
54-1613 & (-1) & N & TO & 1.13 & 1.9 & 3.5 & 1.1 & 0.00$^{+0.01}_{-0.05}$ & 0 &    HR\\  
54-1547 & -33,(-34) & B & CTTS & 0.71 & 1.6 & 9.0 & 1.1 & 0.21$^{+0.15}_{-0.08}$ & 0.21 &   \\   
" & -33,(-34) & B & CTTS & 0.69 & 1.5 & 9.9 & 1.1 & 0.18$^{+0.13}_{-0.06}$ & 0.18 &   \\   
21384350 & (-23) & N & TO & 0.14 & 1.0 & 7.0 & 0.4 & 0.00$^{+0.05}_{-0.01}$ & 0 &    HR \\ 
12-2363 & (-3) & N & N & 0.97 & 2.3 & 1.0 & 0.5 & 0.2$^{+0.6}_{-0.2}$ & 0 &  SB1, HR \\  
12-595 & (-17) & N & TO & 0.35 & 1.2 & 22.9 & 0.9 & 0.6$^{+0.3}_{-0.3}$ & 0 &   scattering? \\  
12-1423 & (-2) & N & N & 0.78 & 1.8 & 4.2 & 0.9 & 0.04$^{+0.03}_{-0.03}$ & 0 &    \\  
12-1010 & -23,(-20) & B & CTTS & 0.19 & 1.2 & 6.0 & 0.4 & 0.04$^{+0.07}_{-0.02}$ & 0.04 &   \\   
12-2098 & (-7) & N & --- & 0.48 & 1.9 & 1.8 & 0.4 & 0.03$^{+0.08}_{-0.02}$ & 0 &   HR \\  
91-506 & -31,(-47) & B & CTTS & 0.70 & 1.7 & 6.5 & 0.9 & 0.09$^{+0.07}_{-0.04}$ & 0.09 &   \\   
12-1617 & -13,(-30) & B & CTTS & 0.67 & 2.0 & 1.6 & 0.5 & 0.16$^{+0.38}_{-0.09}$ & 0.16 & HR \\   
21-840 & (-14) & --- & CTTS & 0.26 & 1.3 & 6.4 & 0.5 & 0.06$^{+0.08}_{-0.03}$ & 0.06 &   \\   
13-1048 & -8,(-7) & B & CTTS & 0.57 & 1.7 & 4.3 & 0.7 & 0.04$^{+0.09}_{-0.03}$ & $<$0.3 &   \\   
13-1250 & -4,(-2) & B & TO & 1.71 & 2.2 & 3.9 & 1.3 & 0.33$^{+0.11}_{-0.10}$ & 0.33 &   \\   
21-563 & (-2) & N & TO & 1.11 & 2.6 & 0.8 & 0.5 & 0.02$^{+0.19}_{-0.03}$ & 0 &  HR \\  
24-542 & (-4) & N: & N & 1.72 & 2.1 & 5.0 & 1.4 & 0.24$^{+0.19}_{-0.09}$ & 0.24 &   now accreting? \\   
24-515 & (-11) & P & TO & 0.34 & 1.4 & 5.6 & 0.6 & 0.10$^{+0.18}_{-0.05}$ & 0.10 &   \\   
21-998 & (-16) & P & CTTS & 0.67 & 1.5 & 9.7 & 1.1 & 0.33$^{+0.19}_{-0.12}$ & 0.33 &  SB1 \\   
21-33 & -51,(-107) & B & CTTS & 0.25 & 1.2 & 9.6 & 0.6 & 0.07$^{+0.11}_{-0.04}$ & 0.07 &   \\   
24-48 & (-1) & N & N & 0.65 & 1.9 & 1.8 & 0.5 & -0.03$^{+0.05}_{-0.02}$ & 0 &   HR \\ 
21-230 & (-3) & N & N & 0.28 & 1.3 & 7.6 & 0.6 & 0.00$^{+0.01}_{-0.01}$ & 0 &   \\  
21-1536 & (-26) & --- & CTTS & 0.23 & 1.1 & 8.8 & 0.6 & 0.18$^{+0.29}_{-0.08}$ & 0.18 &   HR \\  
52-1649 & (-3) & --- & --- & 0.45 & 1.2 & 23.2 & 0.9 & 0.01$^{+0.01}_{-0.01}$ & 0 &   \\  
21-2251 & (-4) & N & N & 0.83 & 2.4 & 0.9 & 0.4 & 0.07$^{+0.26}_{-0.05}$ & 0 &  SB1, HR \\  
21-1586 & (-16) & N & N & 0.30 & 1.1 & 12.3 & 0.8 & 0.01$^{+0.04}_{-0.01}$ & 0 &   HR \\  
53-1762 & (-2) & N & N & 0.25 & 1.1 & 16.1 & 0.7 & 0.01$^{+0.03}_{-0.01}$ & 0  &  \\  
21-1692 & (-5) & N & N & 0.25 & 1.2 & 6.9 & 0.5 & -0.004$^{+0.006}_{-0.003}$ & 0  &   \\  
21-763 & (-3) & N & N & 0.37 & 1.4 & 8.3 & 0.7 &  -0.002$^{+0.015}_{-0.003}$ & 0 &   \\   
21-895a & (-1) & P & TO:$^*$ & 0.97 & 1.8 & 7.7 & 1.1 & 0.21$^{+0.09}_{-0.07}$ & 0.21 &   \\   
21400451 & (-32) & --- & CTTS & 1.34 & 2.1 & 4.9 & 1.2 & 0.71$^{+0.36}_{-0.23}$ & 0.71 & \\   
" & (-32) & --- & CTTS & 1.07 & 1.9 & 6.8 & 1.2 & 0.44$^{+0.21}_{-0.14}$ & 0.44 &   \\   
21-1762 & (-3) & N & N$^*$ & 0.86 & 1.7 & 8.7 & 1.1 & 0.07$^{+0.02}_{-0.02}$ & 0.07/0 &  large hole? \\   
24-1736 & -33,(-61) & B & CTTS & 0.27 & 1.3 & 4.7 & 0.4 & 0.3$^{+0.5}_{-0.1}$ & 0.3 &   \\   
24-1796 & -73,(-124) & B & TOe8 & 0.49 & 1.4 & 9.5 & 0.9 & 0.6$^{+0.4}_{-0.3}$ & 0.6 &  SB2: \\   
21-2006 & (-2) & --- & CTTS & 0.81 & 1.6 & 11.5 & 1.1 & 0.12$^{+0.09}_{-0.05}$ & 0.12 &   \\   
" & (-2) & --- & CTTS & 0.68 & 1.5 & 17.6 & 1.0 & 0.13$^{+0.09}_{-0.05}$ & 0.13 & 4d \\   
22-2651 & -48,(-34) & B & CTTS & 0.26 & 1.3 & 21.2 & 0.8 & 0.05$^{+0.08}_{-0.02}$ & 0.05 &   scattering? \\   
22-1418 & (-2) & B & CTTS & 0.88 & 2.4 & 0.9 & 0.4 & 0.02$^{+0.09}_{-0.02}$ & $<$0.3 &   HR \\  
23-405 & -9,(-14) & B & CTTS & 1.06 & 1.9 & 8.1 & 1.2 & 0.22$^{+0.14}_{-0.08}$ & 0.22 &  \\   
23-570 & -18,(-47) & B & CTTS & 0.92 & 1.8 & 6.4 & 1.1 & 0.7$^{+0.5}_{-0.3}$ & 0.7 &   \\   
\\
\\
\\
\\
\\
\\
\\
\\
{\bf NGC~7160}\\                       
01-580 & (-96) & --- & CTTS & 1.01 & 1.7 & 11.4 & 1.2 & 4$^{+3}_{-2}$ & 4 & \\    
\enddata
\tablecomments{Accretion and stellar parameters. The H$\alpha$ EW has been
compiled from Sicilia-Aguilar et al. (2005b, 2006b, 2008a). 
The accretion rates are given in units of 10$^{-8}$ M$_\odot$ yr$^{-1}$.
It is negative for emission
lines, and the values between parentheses correspond to low-resolution spectroscopy.
The H$\alpha$ profile is marked as `broad' (B),
`probably broad' (P), and `narrow' (N; Sicilia-Aguilar et al. 2006b). The disk type is classified as `CTTS'
for normal classical TTS, `TO' (or `TOe8' if only a small excess at 8$\mu$m is present) 
for transition objects, and `N' for objects without a detectable excess. Uncertain values
are marked with `:'. Objects that have no exess down to 8$\mu$m and have not been
detected at 24$\mu$m, but display some signs of accretion, are marked with $^*$ and
classified as potential TO with large holes.
\.{M}$_U$ indicates the accretion directly
measured from the U band photometry, \.{M}$_{fin}$ is the final value of the accretion rate
after requiring a detection probability $>$99\% (based on 5000 simulations)
and H$\alpha$ (see text). Accretion rates are marked as ``0" when the simulated data is consistent with zero 
and there is no further evidence of accretion (see text). If the photometry is consistent
with no accretion, but H$\alpha$ suggest that the object is accreting, we consider it as an upper limit. 
Whenever the measured U band luminosity was found to be below the expected photospheric luminosity
in U (L$_U$ is negative), we mark it as a negative accretion rate (corresponding to the
accretion rate resulting from the absolute value of L$_U$). 
The errors in the luminosity are typically low ($\sim$ few percent). Radii, masses, and
ages are derived from the V vs. V-I diagram (or the HR diagram, see text, 
marked with HR in the ``Comments" column), using the Siess et al. (2000) isochrones.
For K and M stars, the typical errors in these quantities are $\sim$10\%, although
the age could be more affected by the presence of unresolved binaries. The age of
early K and G-type stars is far more uncertain due to general problems establishing
the birthline (Hartmann 2003); such objects are marked as ``G-age''.
Binary information for spectroscopic binaries (unresolved, similar mass, single-lined [SB1] or
double-lined [SB2] spectroscopic binaries) is taken from Sicilia-Aguilar et al. (2006b).
}
\end{deluxetable}

\begin{deluxetable}{lcccl}
\tabletypesize{\scriptsize}
\tablenum{4}
\tablecolumns{3} 
\tablewidth{0pc} 
\tablecaption{Significance of the trends in \.{M} versus disk structure, age and mass \label{correlation-table}} 
\tablehead{
 \colhead{Trend} & \colhead{r} & \colhead{p}& \colhead{Confidence interval}  & \colhead{Comments}  } 
\startdata
\.{M} vs. $\alpha$(H-4.5) 	& 	0.10	&  0.52	& -0.26,0.44	&  No significant correlation  \\
\.{M} vs. $\alpha$(H-8.0) 	&	0.21	&  0.16	& -0.15,0.53	&  No significant correlation \\
\.{M} vs. $\alpha$(3.6-5.8)	&  	0.38	&  0.02 &  0.03,0.64	&  Moderate correlation \\
\.{M} vs. $\alpha$(5.8-8.0) 	& 	0.02	&  0.87	& -0.33,0.38	&  No significant correlation  \\
\.{M} vs. $\alpha$(H-24) 	&	-0.01	& 0.97	&  -0.38,0.37	& No significant correlation (survey not complete at 24$\mu$m) \\
\.{M} vs. $\alpha$(8.0-24) 	&  	-0.26	& 0.12	&  -0.58,0.13	& Weak potential anticorrelation (survey not complete at 24$\mu$m)  \\
\.{M} vs. age 	(All)	& -0.60 	& $<$0.001	&  -0.70,-0.48	&  Anticorrelation  \\
\.{M} vs. age 	(Cep OB2 only)	& -0.32 	& 0.03	&  -0.61,0.04	&  Moderate anticorrelation  \\
\.{M} vs. M$_*$ 		& 0.32		&  0.03	&  -0.03, 0.59	& Moderate correlation  \\
\enddata
\tablecomments{Spearman's correlation coefficients (r), probability (p) and conficence 
intervals of the different trends explored. A value
of r$\sim$1 indicates a perfect positive correlation, r$\sim$-1 indicates anticorrelation,
r$\sim$0 suggests that the two quantities are uncorrelated. The probability that such
a correlation coefficient is from a sample of random data is indicated by p. }
\end{deluxetable}

\begin{deluxetable}{lccl}
\tabletypesize{\scriptsize}
\tablenum{5}
\tablecolumns{4} 
\tablewidth{0pc} 
\tablecaption{Variable accretion (type II) candidates \label{exors-table}} 
\tablehead{
 \colhead{ID} & \colhead{\.{M}$_{02/03}$ (10$^{-8}~$M$_\odot$~yr$^{-1}$)}  & \colhead{\.{M}$_{07}$ (10$^{-8}~$M$_\odot$~yr$^{-1}$)} & \colhead{Comments}  } 
\startdata
72-875 & $<$0.001 & 0.11$^{+0.23}_{-0.06}$ & SB1: \\
14-141  & $<$0.1 & $<$2.3  & spectroscopically confirmed \\
11-2146 & 16 & 1.0$^{+0.7}_{-0.3}$ & spectroscopically confirmed, type III variable\\ 
11-2031 & 1.6 &  0.3$^{+0.2}_{-0.2}$ & \\
13-1238 & 11 & 0.6$^{+1.1}_{-0.3}$ & \\
82-272 & 24 &  $<$0.9 & SB2, uncertain, type III variable\\
13-277 & $<$30 & 14$^{+6}_{-4}$ & GM Cep, spectroscopically confirmed \\
13-236  &  2 & 0.5$^{+0.3}_{-0.2}$ & \\
13-1250 & 0.1 & 0.33$^{+0.11}_{-0.10}$ & TO \\ 
21-1762 & 0.2 & 0.07$^{+0.02}_{-0.02}$ & large hole candidate \\  
\enddata
\tablecomments{Candidates to have type II variability, related to a variable
accretion rate. }
\end{deluxetable}

\begin{deluxetable}{lccl}
\tabletypesize{\scriptsize}
\tablenum{6}
\tablecolumns{4} 
\tablewidth{0pc} 
\tablecaption{Variable extinction (type III) candidates\label{uxors-table}} 
\tablehead{
 \colhead{ID} & \colhead{A$_V$ (2001)}  & \colhead{A$_V$ (2007)} & \colhead{Comments}  } 
\startdata
11-2146 & 2.58 mag & 1.62$\pm$0.13 mag & type II variable\\
14-287 & 2.20 mag & 1.1$\pm$0.3 mag & disk \\
11-1067 & 1.20 mag & 0.64$\pm$0.12 mag & \\
11-581 & 1.50 mag & 3.19$\pm$0.09 mag & \\
82-272 & 3.58 mag & 3.01$\pm$0.06 mag & disk, SB2, type II candidate \\
12-1955 & 1.42 mag & 0.71$\pm$0.09 mag & \\
12-595 & 1.18 mag & 3.1$\pm$0.4 mag & TO, evidence of scattering \\
21-840 & 2.15 mag & 1.45$\pm$0.04 mag & disk\\
13-1048 & 1.67 mag & 0.71$\pm$0.03 mag & disk \\
21-230 & 1.50 mag & 0.71$\pm$0.03 mag & \\
21-1692 & 1.68 mag & 0.71$\pm$0.03 mag & \\
21-763 & 1.20 mag & 0.61$\pm$0.03 mag & \\
21-1762 & 1.75 mag & 1.57$\pm$0.03 mag & \\
\enddata
\tablecomments{Candidates to have type III variability, related to a variable
extinction, probably caused by circumstellar matter. }
\end{deluxetable}

\end{document}